\documentclass{aa}  
\usepackage{graphicx}
\usepackage{txfonts}
\usepackage{lipsum}
\usepackage{subcaption}         
\usepackage{lscape}            
\usepackage{placeins}           
\usepackage{xcolor}
\usepackage{hyperref}

\begin{document} 

\makeatletter
\let\linenumberpar\relax
\let\linenumbers\relax
\let\runningpagewiselinenumbers\relax
\let\pagewiselinenumbers\relax
\let\resetlinenumber\relax
\let\setrunninglinenumbers\relax
\let\switchlinenumbers\relax
\let\linenumber\relax
\makeatother

   \title{Error dependencies in the space-based CNEOS fireball database}
   \titlerunning{Error in CNEOS fireball database}
   
   \author{E. Peña-Asensio \inst{1}
          \and
           H. Socas-Navarro\inst{2,3}
          \and
          D. Z. Seligman\inst{4,5}
          }
    \institute{
        Department of Aerospace Science and Technology, Politecnico di Milano, Via La Masa 34, 20156 Milano, Italy\\
        \email{eloy.pena@polimi.it, eloy.peas@gmail.com}
        \and
        Instituto de Astrofísica de Canarias, Vía Láctea S/N, 38205 La Laguna, Tenerife, Spain
        \and
        Departamento de Astrofísica, Universidad de La Laguna, 38205 La Laguna, Tenerife, Spain
        \and
        NSF Astronomy and Astrophysics Postdoctoral Fellow
        \and
        Department of Physics and Astronomy, Michigan State University, East Lansing, MI 48824, USA\\
    }
 
  \abstract
   {The CNEOS database offers near-global coverage of fireball events based on U.S. Government sensor detections. The database contributes to investigations of meteoroid impact fluxes. However, the accuracy of these data is not reported.} 
   {We aim to evaluate the reliability of CNEOS-derived ephemerides of fireball events given the absence of the underlying data. In particular, we identify conditions leading to larger or smaller orbital uncertainties and analyze the self-consistency of velocity vectors as a proxy for errors.} 
   {We analyzed 18 events that have both (i) sufficient satellite information to derive orbits and (ii) ground-based observational counterparts. Specifically, we quantified the uncertainties on these ``calibrated events'' using the orbital similarity metric, $D_D$. We also examined the geocentric velocity components imbalance at the population level and identified discriminants that can indicate the accuracy of a given event. } 
   {We identified two distinct groups in the CNEOS database. CNEOS data produces ephemeris determinations with $D_D<0.1$ for fireballs reported either (i) after late 2017 or (ii) with impact energies above 0.45~kt, with 74-78\% of events having $D_D = 0.03 \pm 0.02$ and $\sim$11\% showing $D_D < 0.008$. Our statistical test confirms these two parameters as the only reliable discriminants that, when combined, explain the two accuracy groups. Daylight, the $z$-velocity component, low altitude, duration, and latitude might also indicate errors, although the limited dataset may obscure correlations. No clear discriminants are identified for more restrictive $D_D$ cutoffs. We provide estimates of orbital uncertainties for calibrated events. The hyperbolic fireball subset in the CNEOS database appears as an outlier in the velocity imbalance test.} 
   {Our results confirm that the fidelity of CNEOS fireball data improved significantly from 2018, likely due to the deployment of next-generation space sensors, and show a growing number of high-velocity events. Hyperbolic candidates should be interpreted with caution, as their velocities and inclinations likely reflect measurement errors. Accuracy constraints remain limited by the dataset size, as is evidenced by the lack of statistically significant dependence on duration, preventing strong conclusions from being drawn.}

   \maketitle
%

\section{Introduction} \label{sec:intro}

The database of bright fireball events hosted by the Center for Near-Earth Object Studies (CNEOS)\footnote{\url{https://cneos.jpl.nasa.gov/}} is derived from observations made by U.S. Goverment (USG) space-based sensors. These sensors include classified spy satellites originally developed under the framework of the Treaty on the Non-Proliferation \citep{Tagliaferri1994hdtcconf199T}. These satellites were designed to monitor nuclear explosions and therefore provide near-global coverage of the Earth (60-80\%) \citep{Brown2002Natur420294B}. As a byproduct of their operations, the USG sensors detect fireballs in the silicon bandpass. This is in stark contrast to typical ground-based, fireball-monitoring networks, which are limited to surveying small atmospheric volumes. For instance, the European Fireball Network (EN) monitors only 0.14\% of the Earth's surface \citep{Borovicka2022AA_I}. However, the specifics of the USG space-based instruments, including their sensitivity, data processing pipelines, and measurement uncertainties, are not publicly disclosed by CNEOS. Consequently, while the CNEOS dataset is invaluable for providing broad statistical insights into the frequency and characteristics of meter-sized impactors \citep{Brown2002Natur420294B, Borovicka2009, Brown2013Natur503238B, Farnocchia2017, Palotai2019, Eloy2022AJ16476P, Chow2025116444Icarus}, the lack of transparency regarding the instrumentation and data calibration inhibits its utility for high-precision analyses of individual targets. 

The first step in studying fireballs is to determine their apparent radiant and velocity at the beginning of the atmospheric flight, which are then used to compute the heliocentric orbit. The apparent radiant is defined as the unit vector parallel to the fireball's trajectory as it enters Earth's atmosphere, but with the opposite sign. An accurate assessment of the radiant and initial velocity has been repeatedly demonstrated to be critical for understanding the origin of any given impactor \citep{Egal2017Icar29443E, Hajdukova2019msmechapter, Hajdukova2020PSS19004965H, Hajdukova2020PSS19205060H, Vida2020MNRAS4913996V}. 

Several studies have examined the reliability of the CNEOS fireball data, yielding mixed conclusions. \citet{Devillepoix2019} compared ten fireballs in the CNEOS database that had independent measurements and state that they are generally unreliable for orbit calculations. \citet{Devillepoix2019} identified extreme deviations in USG estimates compared to ground-based values: 28\% in velocity for the Romania and Buzzard Coulee events, and up to 90 degrees in radiant for the Buzzard Coulee and Crawford Bay events. \citet{Eloy2022AJ16476P} later identified a typographical error in the 2008 TC3 fireball data, the correction of which led to a substantial improvement in the agreement for that specific event. \citet{Eloy2024Icar40815844P} concluded, after comparing 17 events with independent ground-based measurements, that approximately 65\% of CNEOS fireball measurements are in relative agreement with ground-based observations. \citet{BrownBorovicka2023} reported that radiant position errors frequently exceeded 10$^\circ$ for the same 17 events. They also extrapolated velocity uncertainties of up to 10–15~km/s for high-speed events, based on the assumption that velocity uncertainty positively correlates with speed. Finally, using a statistical approach, CNEOS data were found to be insufficiently accurate for \citet{Hajdukova2024AA691A8H} to determine the fireball's origin. USG energy measurements of fireballs align within a factor of two for well-documented events observed by GOES satellites \citep{Wisniewski2024Icar41716118W}. 

In this work, we present a comparative analysis of the orbital elements for the 18 CNEOS events with independent ground-based measurements to better constrain the associated uncertainties. These are the only 18 fireball events to date that have independently determined orbits among the 329 fireballs with velocity vector information in the CNEOS database (1007 in total). We adopt the well-established and standard ecliptic J2000 reference frame in order to ensure consistency in our comparisons. This choice reduces potential errors as the literature exhibits inconsistencies in parameter conventions, with multiple definitions often used for the same quantities. For example, some authors do not define azimuth with zero to the north. Velocity may be reported as the corrected pre-atmospheric value, at the onset of the luminous phase, at the fireball’s energy peak, or as an average over its atmospheric flight. Similarly, the atmospheric entry angle is referenced to both the local horizontal and the zenith.

The subset of CNEOS calibrated events has been observed in prior work to comprise two distinct populations based on their errors. Around three quarters of the events exhibit lower $D_D$ values, while the remaining group has significantly larger $D_D$. We performed statistical tests on this set to identify conditions associated with better or worse measurements; we refer to these conditions as discriminants for the remainder of this paper. For example, impact velocity has been widely regarded in the literature as a discriminant because higher-velocity events tend to have larger measurement errors \citep{BrownBorovicka2023, Desch2023arXiv231107699D, Fernando2024GeoJItmp181F, Hajdukova2024AA691A8H}. Although this assumption is valid for ground-based observations and likely applicable to space-based systems, it seems not to be supported by the current CNEOS database \citep{SocasNavarro2024arXiv240517219S}. To refine these insights, we systematically tested all parameters reported in the CNEOS database to identify relationships that explain discrepancies between low- and high-$D_D$ measurements. Hereinafter, we use the terms low-$D_D$ and high-$D_D$ purely for comparison between the two groups—equivalent to referring to lower and higher $D_D$ values. This approach aims to help researchers assess the reliability of events in the database, particularly in cases in which ground-based observations are unavailable.

Additionally, we investigate the reliability of hyperbolic events reported in the CNEOS database based on a velocity component balance analysis. The velocity vectors of Earth impactors are typically described in a geocentric reference coordinate system. The distribution of approach directions of fireballs is random and therefore geocentric fireball velocity components should exhibit a Gaussian distribution. Therefore, there should not be a preferred direction in geocentric fireball velocity components. The orbit of  an impactor with one velocity component significantly higher than its other two would have to align precisely with that randomly chosen axis \citep{Hajdukova2024AA691A8H}. While such alignments can occur by chance for isolated events, it is suspicious if they appear repeatedly within a particular group of measurements. For this reason, consistent imbalances in the velocity components along different axes indicate potential measurement errors. 

This paper is organized as follows. In Section \ref{sec:orbit_error} we outline the methodology for determining heliocentric orbits using data from the CNEOS database, and introduce the $D_D$ criterion as a measure of orbital similarity and describe its application to CNEOS data. In Section \ref{sec:comparisons} we present our comparative analysis of calibrated fireballs and determine effective discriminants to distinguish between low- and high-$D_D$ events. In Section \ref{sec:comp_vel_comp} we introduce a statistical test to analyze the self-consistency of the velocity vectors. In Section \ref{sec:hyp}, we assess the reliability of hyperbolic fireballs in the CNEOS database. Finally, in Section \ref{sec:conclusions} we summarize our findings and discuss their implications for the study of fireballs and potential interstellar impactors. 

\section{Orbit determination and error quantification}\label{sec:orbit_error}

We are primarily motivated to identify conditions under which the observations reported in the CNEOS database exhibit reliable uncertainties. It appears that CNEOS events fall into two distinct subsets: fireballs with larger and smaller uncertainties. In order to quantify the orbital errors, we determine the orbits (\S\ref{sec:orbit}) and implement an objective criterion that is commonly used to determine meteor associations or, equivalently, orbit similarities (\S\ref{sec:error}).

\subsection{Computing ephemerides of fireballs from CNEOS data} \label{sec:orbit}

In this subsection, we describe our methodology to determine heliocentric orbital elements based on the data provided by CNEOS. We first calculate the apparent radiant using the impact velocity vector, $\mathbf{V}$. This vector is defined in the Earth-centered, Earth-fixed (ECEF) coordinate system by CNEOS. In this coordinate system, the $z$ axis aligns with the Earth's rotation axis toward the celestial north pole, the $x$ axis lies in the equatorial plane toward the prime meridian, and the $y$ axis completes the right-handed system. This frame is amenable to the straightforward calculation of parameters that characterize an impactor's motion relative to the surface of the Earth. Examples of these parameters are the atmospheric entry angle (relative to the local horizontal) and azimuth (measured clockwise from true north).

The apparent atmospheric entry direction was calculated with two successive linear transformations applied to $\mathbf{V}=[V_x, V_y, V_z]^T$, which allowed us to compute the local azimuth and elevation (often called slope). The initial transformation is a clockwise (negative) rotation within the $xy$ plane by an angle, $\lambda_e$, defining the longitude at maximal fireball energy\footnote{The longitude of maximal fireball energy is reported by CNEOS.}. The next linear transformation is a clockwise (negative) rotation in the $xz$ plane by the colatitude, defined as $\theta_e = 90^\circ - \phi_e$, where $\phi_e$ is the energy peak latitude. These two transformations in the above order provide the velocity vector in local coordinates:

\begin{equation}
\mathbf{V_{e}}  = 
\begin{bmatrix}
\cos(-\theta_e) & 0 & \sin(-\theta_e) \\
0 & 1 & 0 \\
-\sin(-\theta_e) & 0 & \cos(-\theta_e)
\end{bmatrix}
\cdot
\begin{bmatrix}
\cos(-\lambda_e) & -\sin(-\lambda_e) & 0 \\
\sin(-\lambda_e) & \cos(-\lambda_e) & 0 \\
0 & 0 & 1
\end{bmatrix}
\cdot
\begin{bmatrix}
V_x \\
V_y \\
V_z
\end{bmatrix}.
	\label{eq:v_e}
\end{equation}

In Eq.~\ref{eq:v_e}, the first and second matrices denote the standard rotation matrices about the $z$ axis and $y$ axis by angles $\theta_e$ and $\lambda_e$, respectively. We next calculate the heliocentric orbit of the impactor using the CNEOS provided pre-atmospheric velocity vector, which is then corrected for Earth's rotation and gravitational effects through the zenith attraction method \citep{Whipple1957SCoA1183W, Ceplecha1987BAICz38222C}. The implementation of this methodology is similar to \citet{Eloy2022AJ16476P} and is described with more details in \citet{Eloy2021MNRAS5044829P}. We calculated the ecliptic osculating heliocentric orbital elements by considering the Earth's precise position and distance relative to the Sun. Specifically we calculated the inclination ($i$), eccentricity ($e$), longitude of the ascending node ($\Omega$), perihelion distance ($q$), and argument of perihelion ($\omega$). These parameters provide utility because they (i) characterize individual trajectories and (ii) enable comparative analyses and specifically identification of potential relationships between individual trajectories.

\subsection{Orbital dissimilarity as a proposed criterion for error quantification} \label{sec:error}

In this section, we describe the methodology used to quantify the uncertainties in the CNEOS database. Specifically, we examine fireballs that have been independently observed by ground-based cameras. The orbits provided by these measurements are assumed to be the ground truth. 

We implemented an orbital similarity metric to assess the accuracy of the CNEOS data. Early efforts to quantify the similarity between heliocentric paths of fireball trajectories introduced what is now known as the $D$-criterion. The first of these metrics --- introduced by \citet{Southworth1963SCoA7261S} --- is the $D_{SH}$ criterion which measures orbital dissimilarity. Subsequent advancements refined these criteria to enhance their discriminative power. For instance, \citet{Drummond1981Icar45545D} introduced the $D_D$ criterion as a modification to the $D_{SH}$ metric. $D_D$ incorporates normalized weights for eccentricity and perihelion distance in order to ensure that all terms contribute proportionally to the overall measure. This formulation also incorporates the angle between perihelion points ($\theta_{B A}$), defined as the combined contribution of ecliptic longitude and latitude, and the angle between the inclinations of the orbits ($I_{AB}$). The metric is defined as

\begin{equation}
D_{D}^{2}=\left(\frac{e_{B}-e_{A}}{e_{B}+e_{A}}\right)^{2}+\left(\frac{q_{B}-q_{A}}{q_{B}+q_{A}}\right)^{2}+\left(\frac{I_{AB}}{\pi}\right)^{2}
+\left(\frac{e_{B}+e_{A}}{2}\right)^{2}\left(\frac{\theta_{B A}}{\pi}\right)^{2}.
	\label{eq:D_D}
\end{equation}
In Equation~\ref{eq:D_D}, subscripts $A$ and $B$ denote the two orbits that are being compared.

The $D_D$ metric has been widely used to evaluate orbital similarity, particularly in studies linking meteors to (i) meteor showers or (ii) their parent bodies \citep{McIntosh1990Icar86299M, Spurny2003Natur423151S, Williams2004MNRAS3551171W, Wiegert2005Icar179139W, Rozek2011MNRAS412987R, Abedin2015Icar261100A, Olech2016MNRAS461674O, delaFuenteMarcos2018MNRAS4732939D, delaFuenteMarcos2018MNRAS4733434D, Moskovitz2019Icar333165M, DiMare2019LPICo21096007D, Spurny2020MPS55376S, Janches2020ApJ895L25J, Sergienko2020ARep641087S, Sergienko2020ARep64458S, Andrade2023MNRAS5183850A, Koten2023AA675A70K, Eloy2023MNRAS5205173P, Eloy2024Icar40815844P, Eloy2024PSJ5206P, Humpage2024MNRAS5331412H, Shober2024AA686A130S}. The $D_D$ involves selecting a threshold or cutoff to balance the certainty of accepting or rejecting association hypotheses, minimizing both false positives and missed detections. \citet{Drummond1981Icar45545D} originally proposed a threshold of $D_D = 0.105$ to determine whether two orbits are sufficiently similar to suggest a common origin. However, various authors have employed different thresholds over time that we itemize in Table~\ref{tab:DD_thresholds}. We emphasize that these thresholds were not designed for pairwise orbit comparison (as we are doing here), but rather for evaluating the statistical likelihood of association against a background population.

\begin{table}[h]
\centering
\caption{$D_D$ thresholds applied in various literature studies to assess the similarity of meteoroid stream member orbits.}
\begin{tabular}{ll}
\hline
Reference & cutoff \\
\hline
\citet{Jopek2017PSS14343J} & 0.008 \\
\citet{Brown2016} & 0.05 \\
\citet{Bischoff2024MPS592660B} & 0.05 \\
\citet{Matlovic2020AA636A122M} & 0.06$^\ast$ \\
\citet{Eloy2024AdSpR741073P} & 0.083 \\
\citet{Brown2010Icar20766B} & 0.1 \\
\citet{LePichon2008MPS431797L} & 0.1 \\
\citet{Fu2005Icar178434F} & 0.1 \\
\citet{Drummond1981Icar45545D} & 0.105 \\
\citet{Wu1992MNRAS259617W} & 0.105 \\
\citet{Williams1993MNRAS262231W} & 0.105 \\
\citet{Durisova2024MNRAS5353661D} & 0.105 \\
\citet{Halliday1990Metic2593H} & 0.105 \\
\citet{Koten2006MNRAS3661367K} & 0.105$^\ast$ \\
\citet{Drummond2000Icar146453D} & 0.115 \\
\citet{Sokolova2013AdSpR521217S} & $<$0.2 \\
\hline
\end{tabular}
\tablefoot{$^\ast$Cases where $D_D$ was used alongside additional criteria.}
\label{tab:DD_thresholds}
\end{table}

In this study, we use the $D_D$ criterion to quantify uncertainties in orbital determinations derived from CNEOS data. The CNEOS-derived orbits are compared with independently calculated ground-based orbits, the latter of which are assumed to be accurate. In the remainder of this manuscript we refer to these 18 events with ground-based counterparts as calibrated fireballs. The $D_D$ values for each calibrated fireball enable determination of whether the discrepancy between CNEOS and ground-based data falls below established statistical cutoffs, offering an indicator of the goodness of the observation.

The determination of appropriate $D_D$ thresholds has been extensively refined to account for factors beyond a fixed value. \citet{Nesluvsan1995EMP68427N} introduced the importance of adjusting this threshold based on the background distribution in the specific phase space of each meteoroid stream. For instance, \citet{Jopek1997AA320631J} emphasized that appropriate thresholds should depend on the size of the dataset considered; for example, larger datasets may necessitate stricter orbital-similarity criteria to reduce false positives. \citet{Galligan2001MNRAS327623G} proposed inclination-dependent thresholds: specifically $D_D=0.06$ for $i<10^\circ$, $D_D=0.11$ for $10^\circ<i<90^\circ$, and $D_D=0.18$ for $i>90^\circ$ (retrograde). Near-ecliptic orbits, due to their higher meteoroid density and reduced differentiation by inclination, require lower thresholds to limit contamination by sporadic meteors. Additionally, \citet{Moorhead2016MNRAS4554329M} argued that the threshold should be adapted based on the relative meteor shower and sporadic background magnitudes in order to (i) ensure sensitivity and (ii) minimize contamination. 

\citet{Jopek2017PSS14343J} proposed a statistical approach to define similarity thresholds between orbits based on a proof-of-concept implementation on synthetic data. Their Eq.~16 yields a cutoff value of $\sim$0.008 for a sample size of 800, indicating that two orbits below this threshold can be considered members of the same meteor shower with 99\% confidence. We aim to use the $D_D$ criterion as a metric to assess the observational quality of a fireball, which is distinct from their methodology. Applying their Eq.~16 to a sample size of 2 would yield a cutoff of 0.092. However, this value lies well outside the empirical range for which the formula was calibrated.

We use a cutoff value of 0.1 as a qualitative boundary proxy between low- and high-$D_D$ fireballs. Indeed, our discriminant analysis (\S\ref{sec:discriminants}) shows that, 0.1 is the only threshold that yields statistically significant separation between orbits, while also producing physically meaningful groupings. We note that 0.1 is adopted only for the uncertainty-matching experiment and is not proposed as a general search limit. Furthermore, our approach does not aim to classify orbits as ``good" or ``bad" in an absolute sense. Instead, we quantify the typical dissimilarity with respect to independently measured reference events using the $D_D$ metric. The thresholds applied (0.008, 0.05, and 0.1) serve only as heuristic values to assess how the dataset partitions into subsets of lower and higher dissimilarity, with the latter reflecting larger uncertainties in the derived orbits.

\section{Ground-based sensor measurements versus USG ones} \label{sec:comparisons}

In this section, we evaluate the consistency between ground-based observations and USG sensor measurements, focusing on orbital elements (\S\ref{sec:comp_orbits}), while in Section \ref{sec:discriminants}, we statistically identify discriminants that can indicate the accuracy of a given event. We compare the orbits of 18 calibrated fireballs using data from the CNEOS database alongside corresponding ground-based measurements. We aim to (i) evaluate the consistency between these two sources and (ii) identify thresholds that distinguish between higher and lower reliability in orbit determination.

\subsection{Comparing orbits}\label{sec:comp_orbits}

In Table~\ref{tab:calibrated_data} we summarize data from the CNEOS database for the 18 orbit-calibrated fireballs. In Table~\ref{tab:calibrated_orbits} we show the orbital elements for each fireball both (i) calculated here with CNEOS data (using the methodology outlined in Section \ref{sec:orbit}) and (ii) derived from ground-based measurements with the corresponding references. The Crawford Bay fireball (2017-09-05 05:11:27) was excluded from the analysis due to insufficient data, as its information is limited to a conference abstract and lacks a determined orbit \citep{Hildebrand2018LPI493006H}. 

We also report the $D_D$ value between CNEOS-derived and ground-based orbits for each calibrated fireball calculated using Equation~\ref{eq:D_D} in Table~\ref{tab:calibrated_orbits}. High-$D_D$ orbits are observed for fireballs recorded (i) before mid-2015 or mid-2017 and (ii) with total impact energies below 0.41–0.44~kt (see Figure~\ref{fig:DD_vs_year_lat_energy_vel} in the appendix). In the remainder of this paper we use these cutoffs and the $D_D=0.1$ threshold to refer to high-$D_D$ versus low-$D_D$ as a proxy of the measurement errors (see below for numerical values). The energy and year thresholds are defined by the transition boundaries, with the first value corresponding to the last event meeting the conditions and the second to the first event that does not. 

\begin{figure}
\centering
\includegraphics[width=\columnwidth]{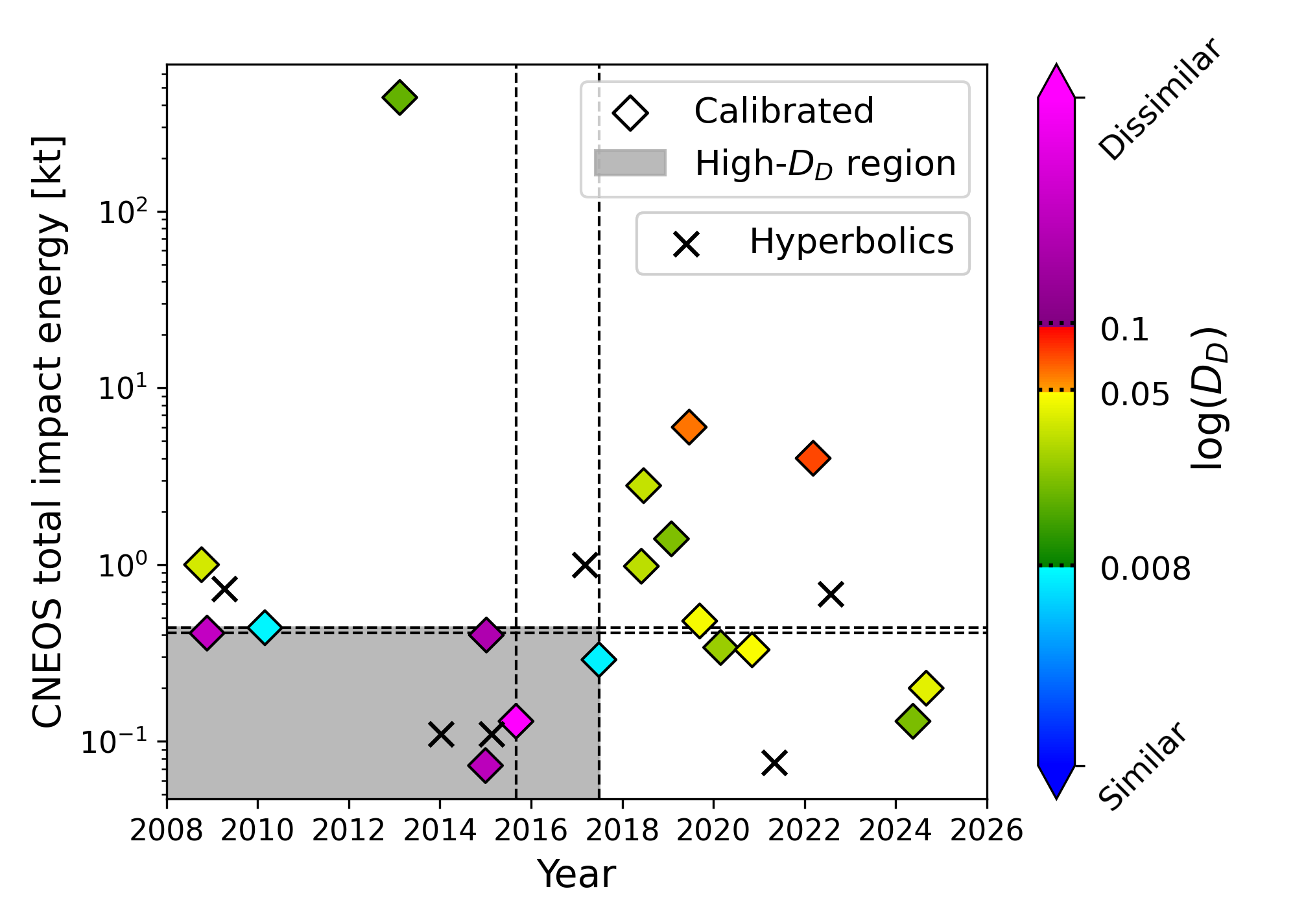}
\caption{Total impact energy provided by CNEOS as a function fireball year. Hyperbolic and calibrated (with independently determined orbits from ground-based observations) events are indicated with diamonds and crosses, respectively. The color of calibrated event markers corresponds to the $D_D$ value between CNEOS-derived and ground-based orbits. Selected $D_D$ thresholds are indicated by a dotted line on the colorbar, delimiting each sub-colormap segment. The gray area highlights the parameter space that results in high-$D_D$ orbits for the calibrated events (based on the criteria $D_D<0.1$). Two vertical and two horizontal dashed lines define the possible boundaries of the high-$D_D$ orbit region. 2008 TC3 has been sign-corrected following \citet{Eloy2022AJ16476P}.}
\label{fig:energy_year_DD}
\end{figure}

Motivated by this stronger dependence of $D_D$ on time and impact energy, we systematically identify energies and fireball occurrence years that produce low- or high-$D_D$. In Figure~\ref{fig:energy_year_DD} we highlight regions of impact energy-year parameter space where high-$D_D$ orbits occur. This region is defined by two possible boundaries for both year and energy, resulting in four possible combinations. These boundaries correspond to: (i) the last event satisfying the $D_D<0.1$ and (ii) the first event that does not satisfy this condition. Based on these thresholds, high-$D_D$ orbits are observed for fireballs recorded (i) before mid-2015 or mid-2017 and (ii) with total impact energies below 0.41 or 0.44~kt. 

Events with $D_D$ between these two thresholds ($0.05 < D_D < 0.1$) are defined as ``doubtful.'' In other words, the classification of these intermediate events as low- or high-$D_D$ is sensitive to the cutoff choice. The most doubtful event is 2022 EB5 ($D_D$=0.072), exhibiting the highest energy peak latitude value (70.0$^\circ$) among all fireballs analyzed (Table~\ref{tab:calibrated_data}). 2019 MO is also classified as doubtful with a $D_D = 0.060$. However, its proximity to the lower end of established thresholds (see Table~\ref{tab:DD_thresholds}) implies it may still represent a reasonably reliable orbit determination. It is worth noting the $\sim168^\circ$ and $\sim180^\circ$ differences in the argument of perihelion and the longitude of the ascending node, respectively, for the event 2019 MO (see Table~\ref{tab:calibrated_orbits}). Due to the orbit’s low inclination, these angular shifts combine geometrically in a way that preserves the spatial orientation of the orbit, thus having a limited effect on the $D_D$ value.

\begin{table*}
\centering
\small
\caption{Physical properties of 18 orbit-calibrated fireballs reported in the CNEOS database.}
\setlength{\tabcolsep}{5pt}
\begin{tabular}{lccccccccccc}
\hline
Event & Date (UTC) & Lat. [º] & Lon. [º] & Alt. [km] & E$_i$ [kt] & $V_x$ [km/s] & $V_y$ [km/s] & $V_z$ [km/s] & Slope [º] & Dur. [s] \\
\hline
2008 TC3 & 2008-10-07 02:45:45 & 20.9 & 31.4 & 38.9 & 1 & -9.0 & 9.0 & -3.8$^\ast$ & 20.8 & 7.5 \\
Buzzard Coulee & 2008-11-21 00:26:44 & 53.1 & -109.9 & 28.2 & 0.41 & 3.9 & -4.1 & -11.6 & 66.7 & 5.0 \\
Košice & 2010-02-28 22:24:50 & 48.7 & 21.0 & 37.0 & 0.44 & -11.7 & 2.7 & -9.1 & 59.8 & 4.5 \\
Chelyabinsk & 2013-02-15 03:20:33 & 54.8 & 61.1 & 23.3 & 440 & 12.8 & -13.3 & -2.4 & 18.6 & 17.0 \\
Kalabity & 2015-01-02 13:39:19 & -31.1 & 140.0 & 38.1 & 0.073 & 4.5 & -14.4 & -10.0 & 20.0 & 10.54 \\
Romania & 2015-01-07 01:05:59 & 45.7 & 26.9 & 45.5 & 0.4 & -35.4 & 1.8 & -4.4 & 43.0 & 3.2 \\
Sariçiçek & 2015-09-02 20:10:30 & 39.1 & 40.2 & 39.8 & 0.13 & 10.3 & -12.2 & -18 & 53.4 & 2.5 \\
Baird Bay & 2017-06-30 14:26:45 & -34.3 & 134.5 & 20.0 & 0.29 & 10.9 & -9.7 & 4.2 & 72.0 & 3.3 \\
2018 LA & 2018-06-02 16:44:12 & -21.2 & 23.3 & 28.7 & 0.98 & 0.9 & -16.4 & 3.9 & 25.1 & 11.5 \\
Ozerki & 2018-06-21 01:16:20 & 52.8 & 38.1 & 27.2 & 2.8 & -8.9 & -4.3 & -10.5 & 77.6 & 4.3 \\
Viñales & 2019-02-01 18:17:10 & 22.5 & -83.8 & 23.7 & 1.4 & -2.4 & 13.6 & 8.7 & 31.8 & 3.5 \\
2019 MO & 2019-06-22 21:25:48 & 14.9 & -66.2 & 25.0 & 6 & -13.4 & 6.0 & 2.5 & 27.9 & - \\
Flensburg & 2019-09-12 12:49:48 & 54.5 & 9.2 & 42.0 & 0.48 & -18.1 & -0.4 & 3.7 & 24.7 & 4.55 \\
Novo Mesto & 2020-02-28 09:30:34 & 45.7 & 15.1 & 34.5 & 0.34 & -18.2 & -11.3 & -2.1 & 47.8 & 3.5 \\
Ådalen & 2020-11-07 21:27:04 & 59.8 & 16.8 & 22.3 & 0.33 & -10.8 & 1.2 & -12.7 & 73 & 4.2 \\
2022 EB5 & 2022-03-11 21:22:46 & 70 & -9.1 & 33.3 & 4 & -11.5 & -5.3 & -11.7 & 56.6 & - \\
Iberian & 2024-05-18 22:46:50 & 41 & -8.8 & 74.3 & 0.13 & -26.5 & -24.1 & 18.7 & 10.5 & 14.0 \\
2024 RW1 & 2024-09-04 16:39:32 & 18 & 122.9 & 25.0 & 0.2 & 3.9 & -19.1 & 2.6 & - & - \\
\hline
\end{tabular}
\tablefoot{We list the date of peak brightness, geodetic latitude and longitude, altitude, total impact energy, and pre-impact velocity components in a geocentric Earth-fixed reference frame with axes aligned to Earth's rotation and equatorial plane. The impact angle was derived following Section~\ref{sec:orbit}; the fireball duration is from references cited in Table~\ref{tab:calibrated_orbits}. $^\ast$Sign corrected following \citet{Eloy2022AJ16476P}.}
\label{tab:calibrated_data}
\end{table*}

\begin{table*}
\caption{Heliocentric orbital elements in the ecliptic J2000 reference frame of calibrated fireballs.}
\centering
\begin{tabular}{lcccccccc}
\hline
Event & Date (UTC) & Reference & $a$ [au] & $e$ & $i$ [º] & $\omega$ [º] & $\Omega$ [º] & $D_D$ \\
\hline
2008 TC3$^\ast$ & 2008-10-07 02:45:45 & \citet{Jenniskens2009} & 1.31 & 0.31 & 2.54 & 234.45 & 194.10 & 0.035 \\
 &  & CNEOS & 1.31 & 0.33 & 2.81 & 239.27 & 194.09 & \\
Buzzard Coulee & 2008-11-21 00:26:44 & \citet{Milley2010} & 1.25 & 0.23 & 25.00 & 211.30 & 238.94 & 0.330 \\
 &  & CNEOS & 0.79 & 0.25 & 10.05 & 3.13 & 238.94 & \\
Košice & 2010-02-28 22:24:50 & \citet{Borovicka2013b} & 2.71 & 0.65 & 2.00 & 204.20 & 340.07 & 0.007 \\
 &  & CNEOS & 2.71 & 0.65 & 3.28 & 204.27 & 340.08 & \\
Chelyabinsk & 2013-02-15 03:20:33 & \citet{Borovicka2013a} & 1.72 & 0.57 & 4.98 & 107.67 & 326.46 & 0.016 \\
 &  & CNEOS & 1.73 & 0.56 & 3.84 & 110.07 & 326.42 & \\
Kalabity & 2015-01-02 13:39:19 & \citet{Devillepoix2019} & 1.80 & 0.50 & 8.73 & 219.80 & 281.62 & 0.290 \\
 &  & CNEOS & 9.03 & 0.90 & 7.93 & 207.67 & 281.62 & \\
Romania & 2015-01-07 01:05:59 & \citet{Borovicka2017} & 2.27 & 0.79 & 12.17 & 98.20 & 106.20 & 0.232 \\
 &  & CNEOS & 4.55 & 0.93 & 20.70 & 112.52 & 106.19 & \\
Sariçiçek & 2015-09-02 20:10:30 & \citet{Unsalan2019MPS54953U} & 1.45 & 0.30 & 22.60 & 182.80 & 159.85 & 1.045 \\
 &  & CNEOS & 0.57 & 0.78 & 40.36 & 358.83 & 159.83 & \\
Baird Bay & 2017-06-30 14:26:45 & \citet{Devillepoix2019} & 1.23 & 0.35 & 3.57 & 259.06 & 98.80 & 0.007 \\
 &  & CNEOS & 1.23 & 0.35 & 3.66 & 260.38 & 98.75 & \\
2018 LA & 2018-06-02 16:44:12 & \citet{Jenniskens2021MPS56844J} & 1.38 & 0.43 & 4.30 & 256.05 & 71.87 & 0.030 \\
 &  & CNEOS & 1.30 & 0.41 & 4.65 & 260.39 & 71.85 & \\
Ozerki & 2018-06-21 01:16:20 & \citet{Kartashova2020105034} & 0.84 & 0.20 & 18.44 & 335.29 & 89.66 & 0.032 \\
 &  & CNEOS & 0.85 & 0.21 & 17.29 & 344.93 & 89.40 & \\
Viñales & 2019-02-01 18:17:10 & \citet{Zuluaga2019} & 1.22 & 0.39 & 11.47 & 276.97 & 132.28 & 0.019 \\
 &  & CNEOS & 1.21 & 0.38 & 9.72 & 277.70 & 132.33 & \\
2019 MO & 2019-06-22 21:25:48 & \citet{JPLHorizons2025a} & 2.53 & 0.63 & 1.56 & 216.78 & 91.04 & 0.060 \\
 &  & CNEOS & 2.39 & 0.59 & 0.23 & 24.31 & 270.91 & \\
Flensburg & 2019-09-12 12:49:48 & \citet{Borovicka2021MPS56425B} & 2.82 & 0.70 & 6.82 & 307.25 & 349.21 & 0.045 \\
 &  & CNEOS & 2.40 & 0.64 & 7.02 & 308.43 & 349.21 & \\
Novo Mesto & 2020-02-28 09:30:34 & \citet{Vida2021EPSC15139V} & 1.45 & 0.61 & 8.76 & 82.65 & 338.99 & 0.024 \\
 &  & CNEOS & 1.45 & 0.59 & 7.88 & 85.02 & 338.98 & \\
Ådalen & 2020-11-07 21:27:04 & \citet{Kyrylenko2023ApJ95320K} & 1.90 & 0.53 & 15.22 & 223.91 & 226.19 & 0.047 \\
 &  & CNEOS & 1.73 & 0.48 & 14.08 & 225.07 & 225.65 & \\
2022 EB5 & 2022-03-11 21:22:46 & \citet{JPLHorizons2025b} & 2.83 & 0.69 & 10.42 & 222.42 & 350.99 & 0.072 \\
 &  & CNEOS & 2.22 & 0.60 & 9.42 & 221.84 & 350.98 & \\
Iberian & 2024-05-18 22:46:50 & \citet{EloyMNRASL2024} & 2.43 & 0.95 & 16.36 & 144.64 & 238.13 & 0.019 \\
 &  & CNEOS & 2.29 & 0.95 & 18.20 & 145.39 & 238.13 & \\
2024 RW1 & 2024-09-04 16:39:32 & \citet{JPLHorizons2025c} & 2.51 & 0.71 & 0.53 & 249.62 & 162.46 & 0.040 \\
 &  & CNEOS & 2.15 & 0.65 & 0.67 & 250.17 & 162.29 & \\
\hline
\end{tabular}
\tablefoot{We show orbital elements computed using ground-based observations (first line for each fireball) and derived from CNEOS data (second line for each fireball). The final column shows the orbital dissimilarity calculated using the $D_D$ criterion. $^\ast$Sign corrected following \citet{Eloy2022AJ16476P}.}
\label{tab:calibrated_orbits}
\end{table*}

The Crawford Bay event (not in the calibrated subset) occurred in September 2017 with a total impact energy of 0.13~kt; both characteristics are consistent with the thresholds for high-$D_D$ orbits. However, the heliocentric orbit of this event is not documented in the literature. Its impact energy falls well below the 0.41–0.44~kt threshold for reliable orbits, and its timing aligns with the mid-2017 cutoff. The event's significant offset in radiant position, reported as 92$^\circ$ by \citet{BrownBorovicka2023}, reinforces its classification as a high-$D_D$ measurement. 

We observe small discrepancies in the reported times of several CNEOS calibrated events when compared to prior studies. For instance, the Buzzard Coulee, Košice, Chelyabinsk, and Kalabity fireballs show differences of 4, 3, 12, and 8 seconds, respectively, relative to the data provided by \citet{BrownBorovicka2023, Eloy2024Icar40815844P}. These discrepancies are likely due to updates made to the CNEOS database, which appears to have been revised in 2024\footnote{During the publication process of this work, CNEOS updated the timestamps of half of the hyperbolic events by 1 s. Energies were recalculated to lower values (difference $<$ 0.1~kt). The event of 8 January 2014 also had its reported position corrected by 0.5°.}. We identified no other typos in the sign of the velocity components in any of the CNEOS events besides from 2008 TC3 \citep{Eloy2022AJ16476P}.

Updates to the CNEOS database such as the one in 2024 may reflect broad improvements in the accuracy of CNEOS data. As noted by \citet{Chow2025116444Icarus}, USG events measured after 2018 exhibit reduced differences in radiant and velocity measurements compared to earlier data. This is consistent with our findings: CNEOS data yield orbits with lower $D_D$ values after late 2017 or for energies larger than 0.45~kt. With an average $D_D = 0.03 \pm 0.02$, 78\% of the events (74\% including the Crawford Bay fireball) have orbits with $D_D$ below 0.1. Using a stricter cutoff of $D_D < 0.05$, 67\% (63\% considering the Crawford Bay fireball) of CNEOS events would have low-$D_D$ orbits, with a mean $D_D$ value of $0.03 \pm 0.01$. Using a more restrictive cutoff of $D_D < 0.008$, 11.1\% of the events would have low-$D_D$ (10.5\% if Crawford Bay fireball is considered).

Table~\ref{tab:errors} summarizes the median $D_D$ values and orbital element uncertainties (with standard deviations) for groups classified as having low- and high-$D_D$ orbits, along with the impact velocity and right ascension ($\alpha_g$) and declination ($\delta_g$) of the geocentric radiant. These uncertainties can serve as reference values for studying CNEOS fireballs when no additional information is available, based on the year of occurrence and the total impact energy. The larger uncertainty in the longitude of the ascending node for the low-$D_D$ group arises from the interplay with the argument of perihelion for the near-ecliptic 2019 MO impactor (see Table~\ref{tab:calibrated_orbits}). As a reference, we report the overall statistics of the $D_D$ values computed for the 17 CNEOS–reference orbit pairs: the median is 0.04, the mean is 0.13, and the standard deviation is 0.24. The difference between the mean and median, along with the high dispersion, indicates a non-Gaussian distribution and supports the presence of two distinct subsets within the sample.

\begin{table}
\centering
\caption{Median and (upper and lower) 1$\sigma$ standard deviations of the $D_D$ values and orbital element, velocity, and geocentric radiant errors for the 18 calibrated fireballs from CNEOS.}
\renewcommand{\arraystretch}{1.5} 
\begin{tabular}{lcc}
\hline
 & High-$D_D$ & Low-$D_D$ \\
 & (Yr. < 2018 \& E$_i$<0.45) & (Yr. $\geq$ 2018 | E$_i$$\geq$0.45) \\
\hline
$D_D$ & 0.31$_{-0.05}^{+0.39}$ & 0.03$_{-0.01}^{+0.02}$ \\
$\sigma_a$ [au] & 1.58$_{-0.92}^{+3.29}$ & 0.04$_{-0.04}^{+0.30}$ \\
$\sigma_e$ & 0.27$_{-0.19}^{+0.17}$ & 0.02$_{-0.01}^{+0.04}$ \\
$\sigma_i$ [º] & 11.74$_{-7.26}^{+4.68}$ & 1.07$_{-0.86}^{+0.26}$ \\
$\sigma_\omega$ [º] & 83.07$_{-69.90}^{+81.44}$ & 1.25$_{-0.66}^{+3.55}$ \\
$\sigma_\Omega$ [º] & 0.009$_{-0.005}^{+0.010}$ & 0.028$_{-0.018}^{+0.225}$ \\
$\mathbf{V}$ & 6.05$_{-2.21}^{+1.44}$ & 0.55$_{-0.45}^{+0.37}$  \\
$\alpha_g$ & 67.24$_{-56.20}^{+69.99}$ & 1.35$_{-0.79}^{+1.12}$  \\
$\delta_g$ & 11.71$_{-4.47}^{+8.78}$ & 0.84$_{-0.79}^{+1.00}$  \\
\hline
\end{tabular}
\tablefoot{Low-$D_D$ and high-$D_D$ are well separated by the cutoff value of 0.1, corresponding to thresholds of events recorded after late 2017 or with impact energies exceeding 0.45~kt.}
\label{tab:errors}
\end{table}

\subsection{Discriminants}
\label{sec:discriminants}

It is evident from the previous subsection that a correlation may exist between orbital accuracy and fireball date and energy. There are two straightforward physical interpretations of this: (i) the satellite network was significantly improved after 2017 and (ii) brighter fireballs (higher impact energy) should be more easily detected and tracked by the sensors, regardless of the (undisclosed) technology employed. In this section we apply a statistical test to quantify the fidelity with which these discriminants provide predictive power. We then use this method to investigate the utility that the remaining CNEOS parameters present as discriminants. 

\begin{figure*}
\centering
\includegraphics[width=1\textwidth]{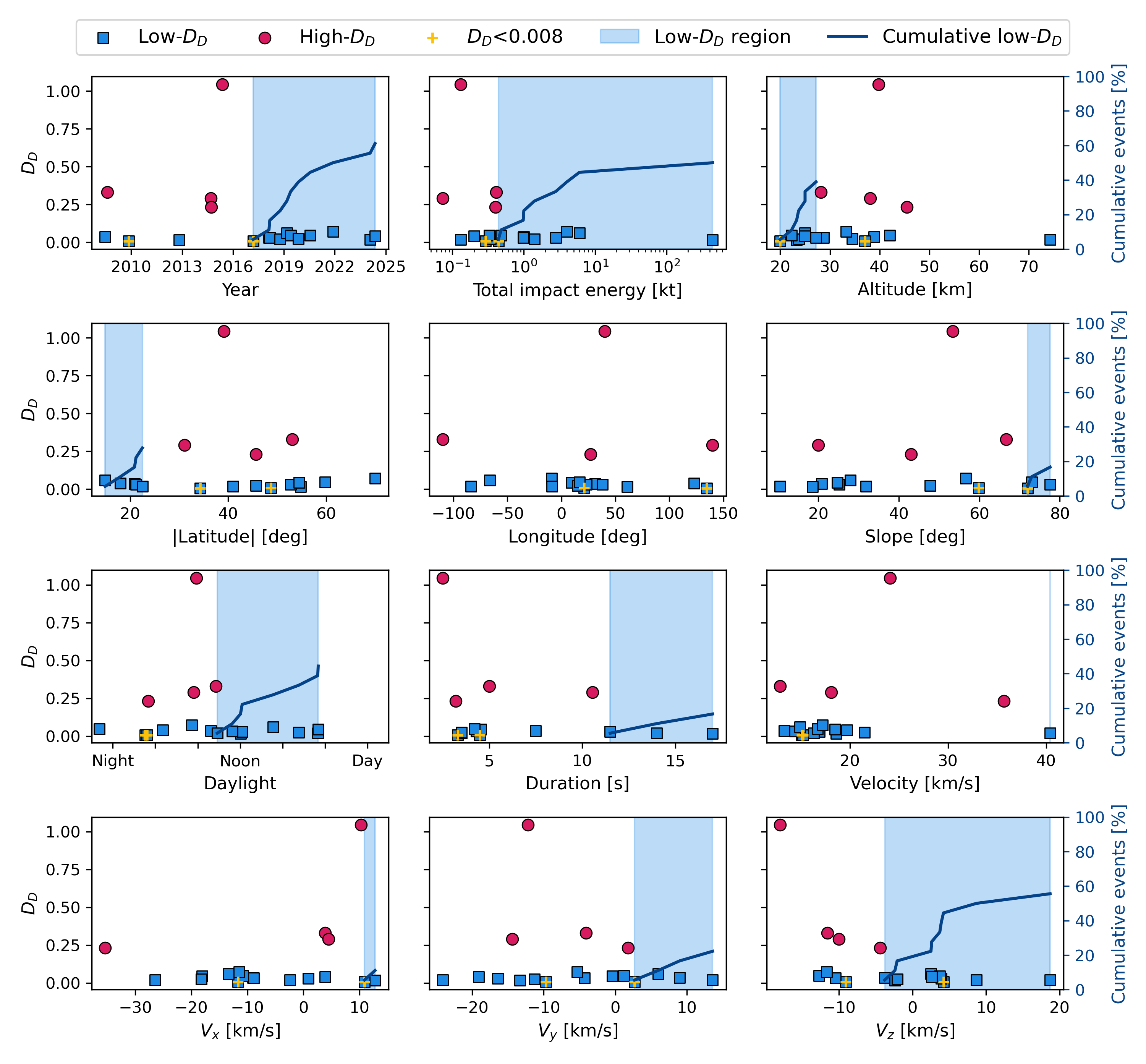}
\caption{Discriminants identified with the statistical test. First-row panels: Date (left), energy (center), and altitude (right). Second-row panels: Absolute latitude (left), longitude (center), atmospheric entry angle (right). Third-row panels: Daylight (left), duration (center), and velocity (right). Fourth-row panels: $x$-velocity component (left), $y$-velocity component (center), and $z$-velocity component (right). The filled light blue region represents consecutive low-$D_D$ events, while the solid line depicts their cumulative percentage, offering a visual indication of the discriminant's effectiveness. Fireballs flagged as low-$D_D$ ($D_D < 0.1$) are denoted by blue squares and high-$D_D$ ($D_D > 0.1$) with red circles. Fireballs with $D_D$<0.008 have a yellow plus symbol.}
\label{fig:discriminants}
\end{figure*}

The fireballs registered in the CNEOS database constitute a series of independent events. Therefore, the accuracy of any given event in the database should be independent of the remaining events. The identification of a fireball event as low-$D_D$ or high-$D_D$ is therefore a Bernoulli experiment. A series of events may be seen as a Poisson process. We investigate the hypothesis that high-$D_D$ events are clustered below or above an event date threshold value, $T$. All high-$D_D$ events occur before 2018 as demonstrated in Figure~\ref{fig:discriminants} (upper left). Therefore define the threshold of $T$ greater than $T=2017.49$~yr as the value that separates the low-$D_D$ and high-$D_D$ groups while also maximizing the number of events in the good group. 

In order to make any claim about the possibility that the event date may be considered a discriminant parameter with threshold $T$, we need to quantify the probability of the null hypothesis. In our case, the null hypothesis is the assumption that the series of consecutive, low-$D_D$ events recorded after the threshold is obtained by chance. The solid blue line in Figure~\ref{fig:discriminants} provides an initial indication of each variable's effectiveness as a discriminant, with a higher probability corresponding to a greater number of events in the light blue filled region.

In a Poisson process, the probability that a series of $n$ consecutive, low-$D_D$ measurements occur by chance is given by $p_B=(p_{B,low})^n$, where $p_{B,low}$ is the overall probability that a given event belongs to the low-$D_D$ group. In our case, the expected value of $p_{B,low}$ is $14/18$. Applying this formula to the series of consecutive low-$D_D$ measurements after $T=2017.49$~yr gives us the resulting probability for the null hypothesis of $0.06$. This low value suggests that the date may be a predictive discriminant with threshold $T=2017.49$~yr, although not with 95\% confidence. Fireballs registered in the database after that date are most likely in the low-$D_D$ group and have their parameters with sufficiently precision to provide an orbit determination with $D_D < 0.1$.

Figures~\ref{fig:energy_year_DD} and ~\ref{fig:discriminants} suggest that the impactor energy might be another predictive discriminant. The statistical test described above provides a low $p_B=0.1$ probability for the corresponding energy null hypothesis ($T=0.44$~kt). Latitude and altitude might also be valid discriminants, although the null hypothesis is not so confidently rejected for these parameters ($p_B=0.28$ for $T=22.5^{\rm o}$ and $p_B=0.17$ for $T=27$~km, respectively). While these $p_B$ values are not conclusive, there are physical reasons why one would expect that these correlations might exist. USG sensors in geostationary orbit have a favorable observing geometry for detecting events near the equator. Moreover, fireballs impacting at lower altitudes should have lower detection errors because they typically reach further atmospheric depths and have longer observable arcs. 

For the longitude parameter, the minimum and maximum values correspond to high-$D_D$ events, so $p_B=1$. It is surprising that velocity is clearly non-discriminant parameter with a $p_B=0.78$ for $T=40.4$~km/s. While some of the highest $D_D$ values are associated with low velocities, numerous low-velocity events also exhibit low $D_D$ values. Additionally, the fastest calibrated fireball--- the Iberian superbolide ---is well aligned with ground-based observations \citep{EloyMNRASL2024}, consistent with the findings of \citet{SocasNavarro2024arXiv240517219S}, although a single case does not necessarily validate a broader trend. However, the Iberian superbolide was long-lasting, with a near-grazing atmospheric entry angle that likely exceptionally enhanced its observability. The $x$- and $y$-velocity components are also non-discriminant parameters. We find a surprising low $p_B$ value for the $V_z$ velocity component ($p_B=0.08$) with threshold $T=-2.4$~km/s. We have not identified an explanation for the correlation between $D_D$ and the sign of $V_z$.

We also considered the possibility that local daytime could be a discriminant because the orbits of daytime fireballs are presumably more difficult to accurately measure. The center panel in Figure~\ref{fig:discriminants} shows the distribution of events as a function of a daylight parameter, which varies linearly local nighttime to local daytime. Contrary to our expectation, the high-$D_D$ group appears clustered in the local nighttime ($p_B=0.13$ for the threshold just before noon). We also tested two parameters that are not provided by CNEOS but we expect to  strongly affect the quality of the observations. Neither the atmospheric entry angle relative to the local horizon ($p_B = 0.37$ for $T = 72^{\rm o}$) nor the fireball duration ($p_B = 0.22$ for $T = 11.5$~s) rejected the null hypothesis with sufficient confidence, although the latter yielded a relatively low $p_B$ value ($\sim$80\% of confidence). This suggests that, while these parameters --- particularly the duration --- are physically meaningful and likely relevant, the current sample size may be insufficient to draw strong conclusions.

For completeness, we repeated the Bernoulli experiment using stricter $D_D$ thresholds of 0.05 and 0.008, effectively redefining the criteria for high- and low-$D_D$ cases. With a cutoff at $D_D < 0.05$, none of the discriminants rejected the null hypothesis. The most significant is $V_z$, with a $p_B = 0.06$, followed by entry angle and altitude (the two of them with $p_B = 0.20$). Only two events satisfied the condition for the more restrictive cutoff at $D_D < 0.008$. This precludes the identification of any discriminant because neither event lies close to the other or near the extremes of any parameter distribution (Figure~\ref{fig:discriminants}). Table~\ref{tab:pb_values} compiles the $p_B$ values and their corresponding threshold for the parameter tested as discriminants.

\begin{table}
    \centering
    \caption{$p_B$ values for different parameters tested as discriminants, under different significance thresholds for three different $D_D$ thresholds.}
    \small
    \begin{tabular}{lcccccc}
        \hline
         $D_D$ cutoff & \multicolumn{2}{c}{0.1} & \multicolumn{2}{c}{0.05} & \multicolumn{2}{c}{0.008} \\
        \hline
         & $T$ & $p_B$ & $T$ & $p_B$ & $T$ & $p_B$ \\
        \hline
        Date & 2017.49~yr & 0.06           & 2024.38 & 0.44       & - & 1 \\
        E$_i$ & 0.44~kt & 0.10             & 2.64~kt & 0.67       & - & 1 \\
        Altitude & 27~km & 0.17            & 23.7~km & 0.20       & - & 1 \\
        Latitude & $|22.5^{\circ}|$ & 0.28 & - & 1                & - & 1 \\
        Longitude & - & 1                  & - & 1                & - & 1 \\
        Slope & 72$^{\circ}$ & 0.37        & 72$^{\circ}$ & 0.20  & - & 1 \\
        Daylight & Pre-noon & 0.13         & Night & 0.30         & - & 1 \\
        Duration & 11.5~s & 0.22           & 11.5~s & 0.30        & - & 1 \\
        Velocity & 40.4~km/s  & 0.78       & 40.4~km/s & 0.67     & - & 1 \\
        $V_x$ & 10.9~km/s & 0.60           & 10.9~km/s & 0.44     & - & 1 \\
        $V_y$ & 2.7~km/s & 0.37            & -16.4~km/s & 0.30    & - & 1 \\
        $V_z$ & -2.4~km/s & 0.08           & 2.6~km/s & 0.06      & - & 1 \\
        \hline
    \end{tabular}
    \label{tab:pb_values}
\end{table}

Our analysis concludes that year and energy are effective discriminants and, when combined, are the only factors that can accurately classify the low-$D_D$ group of all calibrated events. We identified other parameters with suggestive results but lack an intuitive explanation for their significance. We also note that the 2010 Košice and mid-2017 Baird Bay fireballs exhibit the lowest $D_D$ values among all calibrated events. No explanation for this anomaly is evident in the current dataset (see Figure~\ref{fig:discriminants}). The small sample size of calibrated events hinders our ability to provide more conclusive results. As more events are recorded simultaneously by the satellite sensors and ground-based facilities, the number of calibrated bolides will increase and the statistical test will become more reliable. For now, we consider the date and energy as good discriminants, although altitude, latitude, $V_z$, and daylight warrant further investigations.

\section{Analyzing self-consistency of velocity components}\label{sec:comp_vel_comp}

We implemented a methodology to quantify potential differences between subsets of fireballs with the distribution of velocity components inspired by the reasoning described in Section 4.2 of \citet{Hajdukova2024AA691A8H}. Impacts should occur randomly in the geocentric reference frame, thereby producing Gaussian distributions of each individual velocity component. Our methodology hinges on the fact that it is statistically unlikely for any subset distribution to consistently exhibit events with excess in one velocity component. For this same reasoning \citet{Hajdukova2024AA691A8H} stated that the velocity component outlier events in the CNEOS database are caused by measurement errors along one axis\footnote{For more information, see the discussion in the second to last paragraph in the introduction of this paper.}. We introduced a score metric, $\textrm{Score}(\mathbf{V})$, which quantifies the deviation of the individual velocity components from the magnitude of the velocity vector, rather than assessing whether any individual component is extreme relative to the overall population. This metric should determine whether subsets of fireballs exhibit distinctive velocity characteristics. In other words, $\textrm{Score}(\mathbf{V})$ encapsulates how (un)balanced the velocity components are relative to the total velocity and is defined as

\begin{equation}\label{eq:score}
\textrm{Score}(\mathbf{V}) = \sum_{i=x,y,z} \left| \frac{|V_i| - \|\mathbf{V}\|}{\|\mathbf{V}\|} \right|.
\end{equation}

In Equation~\ref{eq:score}, \( \mathbf{V} = (V_x, V_y, V_z)^T \) represents the velocity vector and $\|\mathbf{V}\|$ represents its magnitude. Velocity vectors with high $\textrm{Score}(\mathbf{V})$ have larger imbalances in their components. In other words, any outliers in this metric will exhibit an anomalous excess difference between the values of their components.  \citet{Hajdukova2024AA691A8H}  focused on events with extreme component values relative to the overall population. Our approach is distinct and complementary; we evaluate internal imbalance within each individual event, which does not necessarily imply globally extreme values. Our hypothesis is that this imbalance is caused by measurement errors. To clarify, a randomly chosen subset of events is expected to reflect the population mean. There is no reason for any randomly chosen subset of vectors to all have high $\textrm{Score}(\mathbf{V})$ relative to the overall database regardless of their physical nature.

We next calculated the average score for the subset and compare it to a reference distribution generated through bootstrapping. Bootstrapping is a statistical method that involves generating multiple resampled subsets by randomly drawing data from the original database. This step serves to evaluate whether a subset of fireballs deviates significantly from the population in terms of relative excess in its velocity components. The observed statistic, \( S_{\text{obs}} \), is defined as the mean score across all fireballs in the subset:

\begin{equation}\label{eq:sobserved}
S_{\text{obs}} = \frac{1}{N} \,\bigg[\sum_{n=1}^N \textrm{Score}(\mathbf{V}_n)\bigg]\,.
\end{equation}

In Equation~\ref{eq:sobserved}, \( N \) is the number of events in the subset. The number of bootstrapping iterations of random sampling, \( n_{\text{boot}} = 10,000 \), are performed to construct the reference distribution. The total number of  velocity vectors, \( N \), are sampled without replacement (excluding the subset being evaluated) from the population in each iteration. The mean score for each subset is computed, which produces a distribution of bootstrap statistics \( S_{\text{boot}}^j \). The $p$ value is then calculated as the fraction of bootstrap samples where the mean score, \( S_{\text{boot}}^j \), is greater than or equal to \( S_{\text{obs}} \):

\begin{equation}\label{eq:pvalue}
    p\text{ value} = \frac{1}{n_{\mathrm{boot}}}\,\sum_{j=1}^{n_{\mathrm{boot}}} 
    \mathbf{1}_{\left( S_{\mathrm{boot}}^j \ge S_{\mathrm{obs}} \right)},
\end{equation}

where the indicator function \(\mathbf{1}_{\left( S_{\mathrm{boot}}^j \ge S_{\mathrm{obs}} \right)} \) equals $1$ if $S_{\mathrm{boot}}^j~\ge~S_{\mathrm{obs}}$ and $0$ otherwise.

The assumed null hypothesis is that any randomly drawn subset of velocity vectors should not have an abundance of individual velocity vectors with significant differences between the three components. In other words, a randomly drawn velocity vector should have a low probability of being aligned with one of the (arbitrary) reference coordinate directions. The $p$ value calculated in Equation~\ref{eq:pvalue} represents the likelihood that a subset of vectors would have such anomalous differences. A $p$ value of 1 indicates that the subset's mean score (\( S_{\text{obs}} \)) shows no excess between velocity components. In contrast, a $p$ value near 0 suggests the subset exhibits significant anomalies, enabling the null hypothesis to be rejected. Typically, a $p$ value of \( p = 0.05 \) implies 95\% confidence that the deviation is meaningful and cannot be attributed to random variability.

We analyzed the individual velocity component distributions in order to evaluate the extent to which subsets of fireballs significantly deviate from the overall database. Specifically, we considered the two subsets of calibrated fireballs flagged as (i) low-$D_D$ and (ii) high-$D_D$ based on their orbits. The $p$ values of the two subsets (calculated using Equation~\ref{eq:pvalue}) were (i) $p$ value=0.54 and (ii) $p$ value=0.74, respectively. The $p$ value of the high-$D_D$ subset (ii) is indicative of insufficient statistical evidence to classify them as outliers. Similarly, the $p$ value of the low-$D_D$ subset (i) provides no significant evidence of deviation. While the results do not support a strong statistical imbalance, the high-$D_D$ subset has more velocity component excess than the group with low-$D_D$ orbits.

Figure~\ref{fig:vel_components} in the appendix represents the relationship between total velocity and its velocity components in a geocentric Earth-fixed reference frame within CNEOS database, as in \citet{Hajdukova2024AA691A8H}. In particular, only two events, the Romanian and Iberian fireballs, fall within the tail of the distributions. These two rank among the top 5\% of the highest absolute values for any velocity component. However, only the Romanian event is associated with a poorly measured orbit.

It is important to note that measurement errors can cause deviations in velocity components that still align with the overall database distribution. These errors would produce balanced distributions that appear as having no excess according to our $\textrm{Score}(\mathbf{V})$ metric (Eq.~\ref{eq:score}). A random sample of events with errors following a normal distribution are unlikely to produce outlier imbalances because they tend to cluster around the mean. The 18 calibrated fireballs are effectively a random sample. Therefore, it is likely that they have the typical measurement errors; this is well represented by the $p$ value obtained above. Moreover, some events may deviate from the distribution without necessarily being erroneous. Therefore, while the proposed test is valuable for identifying trends and highlighting potential outliers based on velocity component balance, it should not be relied upon as the sole indicator of orbital accuracy for individual fireballs.

\section{Reliability of hyperbolic fireballs in CNEOS} \label{sec:hyp}

A meteoroid bound to the Solar System typically enters Earth’s atmosphere at speeds between 11.2~km/s and 72~km/s. The lower limit corresponds to Earth’s escape velocity, and the upper limit reflects the sum of the Solar System’s escape velocity at Earth’s orbital distance (42~km/s) and a head-on encounter with Earth’s orbital velocity (30~km/s). If the meteoroid’s heliocentric velocity surpasses 42~km/s, it follows a hyperbolic orbit and may have an interstellar origin.

Scientists have been debating the existence and detections of interstellar meteors since the early 1900s.  \citet{Fisher1928HarCi} reported that nearly 80\% of meteors in Von Niessl and Hoffmeister’s catalog \citep{NiesslHoffmeister1925DKAWW1001V} exhibited hyperbolic trajectories. This initiated a long-standing debate regarding the existence of interstellar meteoroids. This claim was systematically refuted by subsequent studies \citep{Almond1951MNRAS, Opik1956IrAJ, Jacchia1961SCoA, stohl1970BAICz, Hajdukova1993mtpbconf, Hajdukova1994AA288330H}. The vast majority of these hyperbolic orbits result from measurement errors, with only a negligible fraction attributable to Solar System processes such as gravitational scattering by planets \citep{Wiegert2014Icar242112W}.

Modern fireball databases like the EN catalog that are curated with state-of-the-art reduction techniques indicate that $\sim$2\% of impacting meteoroids exhibit hyperbolic orbits within one standard deviation of their measurement uncertainties \citep{Borovicka2022AA_II}. Only two fireballs among these EN hyperbolic events retain hyperbolic classifications within three-sigma confidence limits. The eccentricities in these cases are only marginally above unity (the maximum excess being of $e = 1.048$).

\begin{figure}
\centering
\includegraphics[width=\columnwidth]{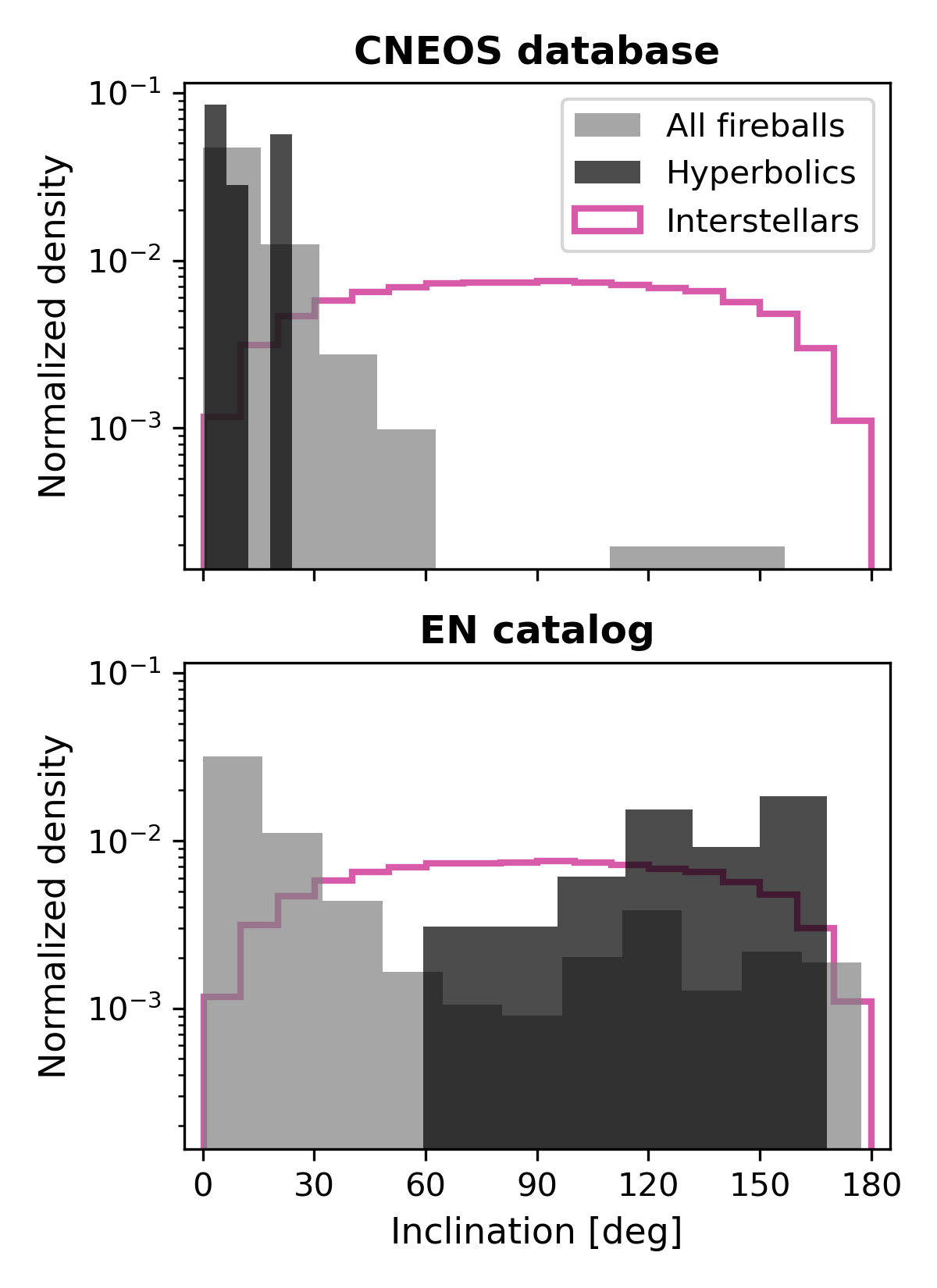}
\caption{Normalized density distributions of orbital inclinations for all fireballs in the CNEOS (top) and EN (bottom) catalogs. An expected orbital inclination for an incoming interstellar population is depicted for reference, following \citet{Marceta2023}. The number of bins was calculated based on Sturges's rule \citep{sturges1926choice}.}
\label{fig:inclinations}
\end{figure}

Moreover, all but one EN hyperbolic meteoroid are in retrograde orbits. This predominance of retrograde orbits is particularly significant as illustrated in Figure~\ref{fig:inclinations}. This retrograde bias has been attributed to the larger relative velocities of retrograde trajectories impacting Earth. These more likely impact head-on, with the higher relative velocities increasing the likelihood of measurement errors \citep{Hajdukova2020PSS19205060H}. Interstellar objects should have kinematic distributions similar to stellar populations. The details of the expected kinematics and sky directions of interstellar objects are complex and depend on the underlying assumptions. For example, different underlying assumptions of the kinematic distributions have resulted in different expectations in terms of the expected distributions of orbital elements and numbers of detections \citep{Engelhardt2014, Cook2016, Seligman2018, Hoover2022PSJ371H, Marceta2023, Marceta2023b}. However, in all of these cases the fraction of prograde to retrograde interstellar objects should be approximately $\sim50:50$ (see Figure 4 in \citet{Marceta2023b}). Therefore, it is safe to say that if these meteoroids were part of a genuine interstellar population, their orbits would statistically show a 50:50 distribution between prograde and retrograde motion. Of course, this is a general consideration about the overall population but it does not say anything about the possibility that any one of those events could be a genuine hyperbolic impactor among many false positives.

\citet{Hajdukova2014MPS4963H} investigated the occurrence of hyperbolic meteoroids and concluded that only an extremely small fraction could credibly be of interstellar origin. They found that most hyperbolic orbits are better explained by inaccuracies in velocity measurements and that  only a small fraction are attributable to Solar System perturbations. Building on this, \citet{Hajdukova2020PSS19004965H} explored how measurement uncertainties impact the determination of meteoroid orbital parameters. Their work demonstrated that even minor errors in velocity or radiant position near the parabolic limit can artificially result in hyperbolic classifications. They emphasized the lack of robust evidence for truly interstellar meteoroids in current datasets.

The discovery of 1I/'Oumuamua in 2017, 2I/Borisov in 2019, and 3I/ATLAS in 2025 marked the first confirmed detections of macroscopic interstellar visitors to our Solar System \citep{Williams2017, borisov2019cbet, Jewitt2023ARAA, Seligman2022ConPh63200S, Jewitt2024, Fitzsimmons2023, Denneau2025, Seligman2025, delaFuenteMarcos2025}. Spacecraft such as Ulysses and Galileo sampled dust particles traversing our Solar System in hyperbolic orbits, indicating a flux of smaller interstellar material \cite{Grun1993Natur362428G, Grun1997Icar129270G, Grun2000JGR10510291G, Landgraf2000JGR10510343L}. Radar detections of faint meteors with hyperbolic orbits have also been reported \citep{Meisel2002ApJ579895M, Weryk2004EMP95221W, Baggaley2000JGR10510353B, Froncisz2020PSS19004980F}. Recently, two bright fireballs identified within the CNEOS space-based database have been highlighted as potential interstellar candidates \citep{Siraj2022ApJ93953S, Siraj2022ApJ941L28S, Eloy2022AJ16476P}: fireballs that occurred on 2014-01-08\footnote{Dates are given in the numerical year–month–day format (YYYY–MM–DD) and, when applicable, with hours–minutes–seconds (hh:mm:ss), to maintain consistency with previous literature and with the format used by CNEOS.} and 2017-03-09. These hyperbolic events have sparked great controversy, particularly regarding claims that remnants of the 2014-01-08 event were recovered from the ocean floor.

Table~\ref{tab:hyperbolics_data_e_i} in the appendix presents the six hyperbolic fireballs identified in the CNEOS database. For each event, the table provides the date of peak brightness, latitude, impact energy (calculated from radiated energy), and the pre-impact velocity components in a geocentric Earth-fixed reference frame. The last two columns present the orbital eccentricity and inclination values as derived in this study.

Figure~\ref{fig:energy_year_DD} shows these hyperbolic events within the context of the calibrated fireballs. 2014-01-08 17:05:34 and the event from 2015-02-17 13:19:50 are situated within the region identified as having high $D_D$. The first one exhibits the highest orbital eccentricity among the hyperbolic fireballs, while the second has the lowest orbital inclination. It is worth noting that a high value of $D_D$ does not necessarily negate the hyperbolic orbit of a particular event, just as a low value of $D_D$ does not confirm the hyperbolic orbit. This is particularly relevant in the case of 2014-01-08 17:05:34 which would require either (i) much larger errors than those observed in the calibrated subset or (ii) very specific combinations of errors in velocity and radiant to turn its orbit into a bound one \citep{SocasNavarro2024arXiv240517219S}. We also highlight the unusually low energy of the 2021-05-06 05:54:27 event, which is substantially lower than the lowest energy among the calibrated fireballs considered with low-$D_D$ orbits. 

Any genuine sample of interstellar fireballs should exhibit a broad range of geocentric speeds as the relative velocities of nearby stars range from 15 to 40~km/s, rather than being confined to exclusively high velocities \citep{Hajdukova2019msmechapter, Marceta2023}. Additionally, there is no physical reason for these fireballs to have significant disparities between their individual velocity components. In other words, the most probable distribution of interstellar objects would have a broad range of speeds with more or less balanced velocity components. It is statistically unlikely for an impactor's motion to align predominantly with one of the randomly oriented axes. By construction, such a scenario would produce an outlier with our velocity component balance metric (see Section \ref{sec:comp_vel_comp}). It is apparent that five out of the six hyperbolic events fall within the tails of the distributions of velocity components in a geocentric Earth-fixed reference frame (Figure~\ref{fig:vel_components} and \citet{Hajdukova2024AA691A8H}). This subset ranks among the top 5\% of the overall database for at least one velocity component. For comparison, ten out of fifteen hyperbolic events in the EN catalog rank within the top 5\%, while fourteen fall within the top 10\%. However, none of these hyperbolic EN fireballs are considered interstellar. For reference and following \citep{Hajdukova2024AA691A8H}, Figure~\ref{fig:vel_components_EN} in the appendix presents the corresponding velocity component representation for the EN catalog.

We calculated the $\textrm{Score}(\mathbf{V})$ metric (Eq.~\ref{eq:score}) for this hyperbolic subset of CNEOS and compared it to the rest of the events. We performed this comparison with a bootstrap approach following the method described in Section \ref{sec:comp_vel_comp}. The hyperbolic subset exhibits a statistically unlikely imbalance in their distribution of velocity components, being distinct from the overall population with a $\sim$90\% of confidence.

If the CNEOS hyperbolic subset represents an outlier, its measurement errors are unlikely to align with the expected values of a normal distribution, potentially limiting the applicability of our $D_D$-derived thresholds. The hyperbolic events 2014-01-08 17:05:34 and 2015-02-17 13:19:50 lie on the high-$D_D$ orbit region of Figure~\ref{fig:energy_year_DD}. Recall that we use the criteria established from the calibrated fireballs to determined if derived orbits have low-$D_D$ or high-$D_D$ (see Section \ref{sec:comp_orbits}). Additionally, we consider the old event occurred on 2009-04-10 18:42:45 as high-$D_D$ despite its energy exceeding 0.45~kt. These threshold were based on the calibrated fireballs which effectively are a random subset. Therefore, the calibrated subset probably does not include outliers, implying that the calibrated fireballs probably do not have unbalanced velocities or hyperbolic orbits. For this reason, we accept the 2009-04-10 18:42:45 fireball as an exception, a conclusion further supported by subsequent results. For the subset of these three events flagged as potentially high-$D_D$ (2014-01-08 17:05:34, 2015-02-17 13:19:50, and 2009-04-10 18:42:45), a $p$ value of 0.003 was obtained, indicating strong evidence that they are statistically distinct from the overall population in terms of the balance of their velocity components. 

\begin{figure}
\centering
\includegraphics[width=\columnwidth]{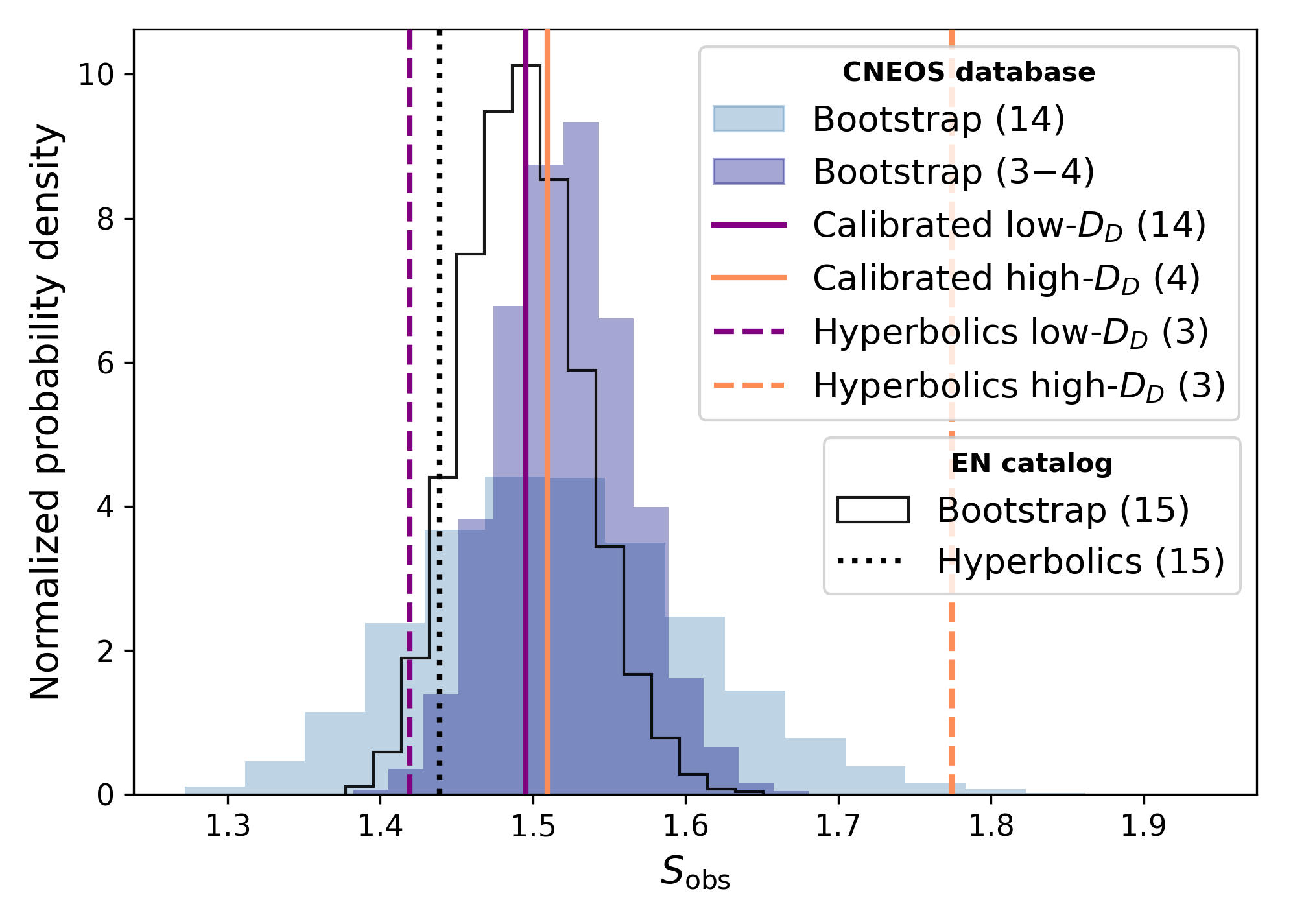}
\caption{Bootstrap results for the $\textrm{Score}(\mathbf{V})$ metric (Eq.~\ref{eq:sobserved}), derived from geocentric velocity distributions. Light blue bars represent the bootstrap distribution generated by randomly selecting 14 samples from the calibrated subset flagged as low-$D_D$ orbits. Darker blue bars represent the average bootstrap distribution for the remaining CNEOS subsets: (i) calibrated flagged as high-$D_D$ orbits (4 samples), (ii) CNEOS hyperbolic flagged as high-$D_D$ (3 samples), and (iii) CNEOS hyperbolic flagged as low-$D_D$ (3 samples). The solid black contour line represents a histogram of bootstrap results from the EN catalog, generated by randomly selecting 15 samples. Sample sizes for subsets and bootstraps are in parentheses in the legend and are paired by their equal values. In all cases, the bootstrap process involved randomly selecting samples and evaluating the score metric 10,000 times. The $\textrm{Score}(\mathbf{V})$ quantifies the relative excess of each velocity component compared to the vector magnitude. Subsets on the right exhibit higher relative differences between velocity components, while those on the left indicate more balanced component values. The mean score is indicated for three groups: (i) calibrated fireball subsets from CNEOS (solid vertical lines), (ii) hyperbolic subsets in CNEOS (dashed lines), and (iii) hyperbolic subsets in the EN catalog (dotted line). The number of bins was calculated based on Sturges's rule \citep{sturges1926choice}.}
\label{fig:bootstrap}
\end{figure}

In contrast, the subset of hyperbolic fireballs comprising of 2017-03-09 04:16:37, 2021-05-06 05:54:27, and 2022-07-28 01:36:08 yields a $p$ value of 0.87, suggesting no excess between components. For comparison, the hyperbolic fireballs in the EN catalog have a $p$ value of 0.93; the EN hyperbolics are not more imbalanced than the entire EN catalog. Figure~\ref{fig:bootstrap} shows the bootstrap mean results for the calibrated and hyperbolic subsets of the CNEOS catalog, with reliability flagged based on our $D_D$-derived criteria. An extreme outlier is evident, corresponding to a CNEOS hyperbolic event flagged as having an high-$D_D$ orbit. For comparison, Figure~\ref{fig:bootstrap} also includes the mean results for hyperbolic events from the EN catalog. Overall, the EN catalog exhibits a slightly more balanced distribution of velocity components, as expected given its high-quality measurements and data processing.

While we cannot entirely rule out the possibility that genuine interstellar fireballs may appear as outliers, or conversely, that measurement errors could mimic such characteristics, a true interstellar subset should not consistently bias the velocity vector distribution. Such biases are more plausibly explained by errors. All $p$ values computed in this work are summarized in Table~\ref{tab:p_values}.

If the CNEOS and EN catalogs were created by the same instruments and observing conditions, we would expect them to have similar overall characteristics. Specifically, if the CNEOS and EN catalogs measured the same population of fireballs with the same precision, there should be $\sim6-7$ hyperbolic events in the CNEOS database. These numbers are calculated because the EN reported $\sim$2\% of impacts with hyperbolics orbits. Another cautionary fact is that the hyperbolic events reported in CNEOS exhibit eccentricities that exceed the maximum value recorded for the most hyperbolic fireball in the EN catalog. Moreover, a significant number of velocity components for the CNEOS hyperbolic fireballs fall within the tails of the distribution. This can be readily explained if these objects are high-velocity Solar System bodies measured with the less accuracy. This was the adopted explanation for the hyperbolic fireballs in the EN catalog \citep{Borovicka2022AA_II}.  

The bold values in Table~\ref{tab:hyperbolics_data_e_i} indicate the unbalanced velocity component. This excess lies in the $xy$ plane in all three cases. This is expected if these hyperbolic impactors are native to our Solar System. This is because they are likely aligned with the ecliptic plane which more closely corresponds to the $xy$ plane of the geocentric reference frame. This trend is also reflected in the mean velocities of the overall population, which are 9.4, 9.7, and 8.5~km/s for $V_x$, $V_y$, and $V_z$, respectively.

\begin{figure*}
\centering
\includegraphics[width=1\textwidth]{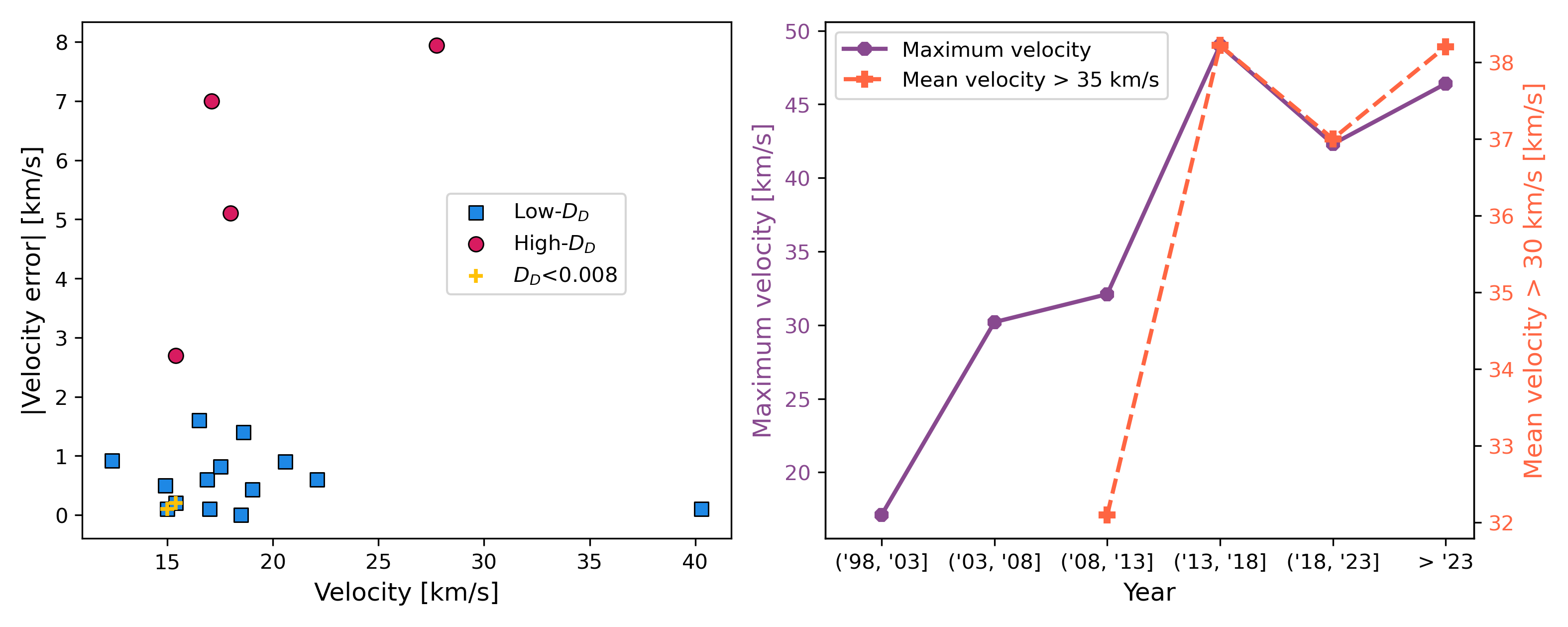}
\caption{Left: Absolute velocity error as a function of velocity for the all calibrated CNEOS events, and for the low-$D_D$ ($D_D < 0.1$) and high-$D_D$ ($D_D > 0.1$) groups. Fireballs with $D_D$<0.008 have a yellow plus symbol. Right: Time evolution of the maximum velocity and the mean velocity of faster events reported by CNEOS.}
\label{fig:year_vel}
\end{figure*}

As is shown in Figure~\ref{fig:inclinations}, the mean inclination for a real population of interstellar objects is around 90$^\circ$. This is far from the case for the hyperbolic orbits in the CNEOS database which are biased toward low-inclination impactors. Indeed, the hyperbolic objects in CNEOS lie at the lower end of the inclination range. In the CNEOS database, the only three retrograde orbits occur in fireballs displaying large deviations in the $V_y$ velocity component, leading to anomalously low-perihelion orbits \citep{Hajdukova2024AA691A8H}. The bias toward prograde, low-inclination impactors in the CNEOS database can also be explained by their velocities, which are slower on average (18.5~km/s), typical of asteroidal fireballs. In contrast, the EN catalog has a higher average velocity (32.9~km/s). The EN catalog is a better representation of the overall near-Earth population because it is sensitive to a wider range of velocities, and consequently to a wider range of inclinations, as it includes cometary fireballs. This apparent bias toward low inclination and velocity in the CNEOS database raises the possibility that USG sensors may be less effective at detecting faster events. This interpretation is tentatively supported by the fact that approximately 20\% of the EN catalog includes fireballs faster than the highest-velocity event recorded in CNEOS. Therefore, if CNEOS is systematically missing the fastest events because they are poorly measured, one would naturally expect the faster events that are reported to exhibit larger velocity uncertainties. However, this trend is not evident in the currently calibrated CNEOS events, as shown in the left panel of Figure~\ref{fig:year_vel}, and was first noted by \citet{SocasNavarro2024arXiv240517219S}. This absence is likely due to dataset limitations, such as selection bias or insufficient sample size, particularly for the low-$D_D$ events, where the apparent trend is strongly influenced by a single data point. Nevertheless, a trend is appreciable only in the high-$D_D$ group. We note that CNEOS has progressively included faster events (see right panel of Figure~\ref{fig:year_vel}) while simultaneously improving accuracy (see top left panel of Figure~\ref{fig:discriminants}). This evolution may complicate the identification of underlying trends, particularly given the small sample size. Nevertheless, we interpret the gradual inclusion of faster events as a reflection of improvements in observational capabilities and/or data reduction methods.

Additionally, almost all hyperbolic fireballs in the EN catalog are retrograde. Moreover, higher inclination fireballs in the EN catalog tend to be correlated with hyperbolic orbits (Figure~\ref{fig:inclinations}). This likely reflects an increased probability of head-on collisions with Earth at near-ecliptic retrograde orbits, which raises the relative velocity and the likelihood of measurement inaccuracies (see the end of Section \ref{sec:intro}). Furthermore, the population of CNEOS hyperbolic fireballs shows an unlikely imbalance in their velocity components. 

Based on our estimated orbital uncertainties (see Table~\ref{tab:errors}), four of the six hyperbolic events could become elliptical within the confidence range of the high-$D_D$ group, assuming similarity to the calibrated subset. The 2014 and 2017 hyperbolic CNEOS events require specific parameter variations to become elliptical, but these variations also fall within the range observed in the calibrated set. Therefore, the CNEOS hyperbolic subset listed in Table~\ref{tab:hyperbolics_data_e_i} is unlikely interstellar; they are more likely high-velocity fireballs resulting from large measurement errors. This is further supported by the fact that all the hyperbolic orbits exhibit prograde motion---see Figure~\ref{fig:inclinations} (top). 

This assessment of the reliability of CNEOS hyperbolic events should be regarded as a statement about the overall population observed in the database and does not refute individual events. Individual analyses would require access to raw data, detailed instrument specifications or independent measurements of these events. 

\begin{table}
\centering
\caption{Results of $p$ values obtained by comparing the average sum of normalized absolute deviations for each subset against the overall database using a bootstrap approach.}
\begin{tabular}{lcc}
\hline
Database & Subset & $p$ value \\
\hline
CNEOS & Calibrated low-$D_D$ & 0.74$\pm$0.44 \\
CNEOS & Calibrated high-$D_D$ & 0.54$\pm$0.50 \\
CNEOS & Hyperbolics & 0.12$\pm$0.32 \\
CNEOS & Hyperbolics low-$D_D$ & 0.87$\pm$0.35 \\
CNEOS & Hyperbolics high-$D_D$ & 0.003$\pm$0.066 \\
EN catalog & Hyperbolics & 0.93$\pm$0.26 \\
\hline
\end{tabular}
\tablefoot{Mean values are accompanied by 1$\sigma$ standard deviations.}
\label{tab:p_values}

\end{table}

\section{Conclusions} \label{sec:conclusions}

Understanding the accuracy of fireball orbital data is essential for assessing meteoroid impact hazards and identifying potential interstellar objects. The CNEOS database, derived from U.S. Government space-based sensors, provides near-global coverage of fireball events, yet the uncertainties in its orbital determinations remain largely uncharacterized due to the classified nature of its instrumentation. In this study, we quantified these uncertainties by systematically evaluating the reliability of CNEOS-derived orbits. To achieve this, we applied the well-established $D_D$ criterion, comparing CNEOS-derived orbital elements to independent ground-based measurements. This allowed us to identify the conditions under which CNEOS orbits have lower or higher $D_D$ values. Additionally, we assessed the balance of geocentric velocity components within the CNEOS database using a bootstrap approach to test for systematic measurement biases, particularly in hyperbolic fireballs.

It is possible that USG sensors correspond to the infrared early warning satellites operated by the U.S. Space Force such as the Space-Based Infrared System (SBIRS). However, neither SBIRS nor its predecessor, the Defense Support Program (DSP), are explicitly mentioned on the CNEOS website. Although this attribution cannot be confirmed, the fact that SBIRS remains the only operational space-based infrared early warning system currently fielded by the United States during decommissioning of DSP provides at least circumstantial evidence to the hypothesis. The transition from DSP to SBIRS began in 2008, with a significant upgrade in operational capacity through SBIRS Block 10 in 2016. Further improvements culminated in the operational acceptance of Block 20 in 2019 \citep{DOTE_SBIRS_2019}. This gradual transition may explain the higher $D_D$ observed prior to this period, as DSP's exclusive use of geostationary satellites limited coverage at higher latitudes. In contrast, SBIRS combines geostationary and highly elliptical orbit sensors, enabling improved spatial coverage and accuracy across a wider range of latitudes \citep{SBIRS_factsheet_2023}. However, we reiterate that this hypothesis is reliant on circumstantial evidence at best.

We conclud that CNEOS events have $D_D<0.1$ orbits either (i) after 2017 or (ii) for fireballs with impact energies exceeding 0.45~kt, which is observed in 74-78\% of the events with an average $D_D = 0.03 \pm 0.02$. Using a more conservative cutoff of $D_D < 0.05$, 63-67\% of CNEOS events have low-$D_D$ orbits, while using $D_D < 0.008$ only $\sim$11\% would have low-$D_D$. Our analysis identifies year and energy as the only reliable parameters that, when combined, effectively classify the accuracy groups of the CNEOS database for $D_D<0.1$. While other parameters, such as daylight and the $z$-velocity component, show potential as reliability indicators, they lack a physical interpretation. Additionally, low altitude, duration, and latitude might correlate with smaller errors. For smaller $D_D$ cutoffs, we did not find either any statistically significant discriminant. However, the limited number of well-characterized events constrains our ability to robustly evaluate parameter dependencies, and likely obscures correlations that may exist in a larger sample.

We provide orbital element (along with radiant and velocity) uncertainties for each group and observe a distinctive improvement in CNEOS data accuracy over time, where faster events have been progressively reported. We conclude that this is likely caused by advancements in the underlying USG sensors. These advancements presumably include the transition from the geostationary DSP satellites to the SBIRS system; the latter incorporates satellites on highly elliptical orbits and provides better monitoring capabilities.

There are six hyperbolic fireballs in the CNEOS database. We find that this hyperbolic subset differs from the overall population with a $\sim$90\% confidence. However, genuine interstellar impactors should not exhibit distinct velocity vector characteristics in a geocentric reference frame. We conclude that the three hyperbolic events before 2018 are extremely likely outliers in terms of the imbalance of their velocity components, including the proposed first interstellar meteor identified by \citet{Siraj2022ApJ93953S}. Therefore, the CNEOS hyperbolic fireball population is unlikely of interstellar origin because:
\begin{enumerate}
    \item They display both an excess in one component of their velocity vector---lying in the tail of the overall fireball distribution \citep{Hajdukova2024AA691A8H}---and a systematic imbalance among the three velocity components.
    \item They are exclusively high-velocity impactors, yet slower interstellar fireballs should also be present \citep{Hajdukova2019msmechapter, Hajdukova2020PSS19205060H, Hajdukova2024AA691A8H}.
    \item They all exhibit prograde motion, but interstellar impactors should have an approximately 50:50 prograde-retrograde distribution \citep{Hajdukova2024AA691A8H}.
    \item Their velocity vector components exhibit an unlikely imbalance.
    \item They differ significantly from hyperbolic fireballs in other reliable databases.
    \item CNEOS fails to report faster fireballs, suggesting potential detection biases against high-velocity events.
    \item >22\% of CNEOS events show significant inaccuracies in a random sample, meaning that if hyperbolics form an outlier subset, their errors may be even larger.
\end{enumerate}

Based on this, CNEOS hyperbolic group is more plausibly explained as mismeasurements of fast impactors. However, without additional information (e.g., raw data, observation geometry, instrument sensitivity), definitive conclusions cannot be drawn for individual events, and true interstellar fireballs may be hidden among uncertainties. 

In summary, while the CNEOS dataset provides valuable global insights into fireball activity, its use for precise orbital determinations requires caution \citep{Devillepoix2019, Eloy2024Icar40815844P, BrownBorovicka2023, Hajdukova2024AA691A8H, Chow2025116444Icarus}, particularly for the older or less energetic impacts. Our work may help establish confidence limits for the events registered in the database. Continued improvements in sensor systems and cross-validation with independent ground-based measurements are essential for advancing our understanding of these phenomena, particularly for identifying potential interstellar fireballs within the database.

\begin{acknowledgements}
      EP-A acknowledges support from the LUMIO project funded by the Agenzia Spaziale Italiana (2024-6-HH.0). DZS is supported by an NSF Astronomy and Astrophysics Postdoctoral Fellowship under award AST-2303553. This research award is partially funded by a generous gift of Charles Simonyi to the NSF Division of Astronomical Sciences. The award is made in recognition of significant contributions to Rubin Observatory’s Legacy Survey of Space and Time. We extend our gratitude to Marco Langbroek for his assistance with information regarding USG sensors. DZS thanks Lia Corrales for helpful conversations. We thank T. Jopek for the clarification on the dependence of the $D_D$ threshold on sample size and the limitations of extrapolating his cutoff formula to individual pairs.
\end{acknowledgements}

\bibliographystyle{aa}
\bibliography{references.bib}

\begin{appendix}
\section{Supporting figures}

\begin{figure}[h]
\centering
\includegraphics[width=\columnwidth]{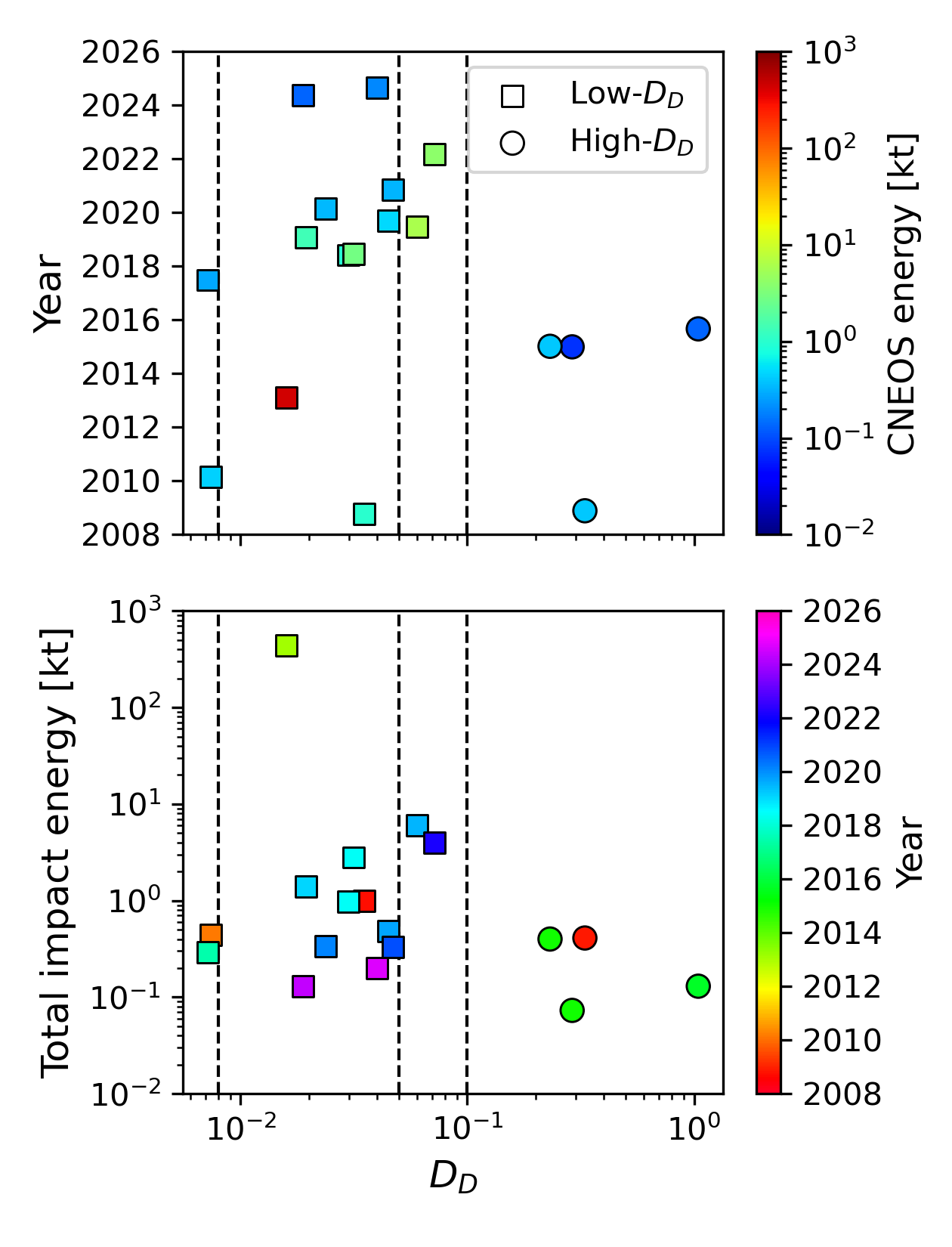}
\caption{Relationship between the $D_D$ values (comparing CNEOS-derived and ground-based orbits) and fireball time (top) and total impact energy (bottom). These values are for the 18 calibrated fireballs, and fireball parameters are those provided by the CNEOS database. Selected $D_D$ thresholds are indicated with dashed vertical lines corresponding to cutoff values of 0.008, 0.05, and 0.1. Fireballs flagged as having low-$D_D$ ($D_D < 0.1$) are denoted by squares and high-$D_D$ ($D_D > 0.1$) with circles. Fireballs in each panel are color-coded according to the $y$ axis variable of the other panel.}
\label{fig:DD_vs_year_lat_energy_vel}
\end{figure}

\begin{figure*}[h]
\centering
\includegraphics[width=1\textwidth]{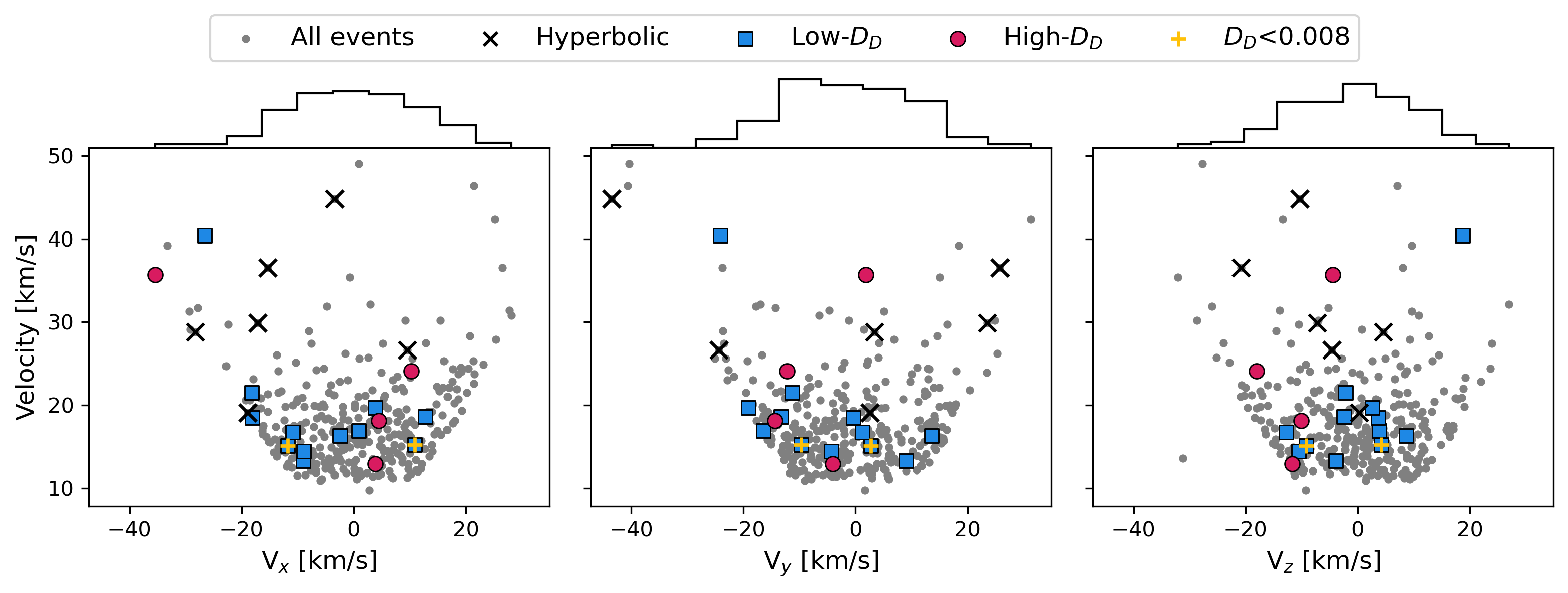}
\caption{Relationship between the total velocity and the pre-impact velocity components in a geocentric Earth-fixed reference frame, as provided by the CNEOS database. Fireballs flagged as having low-$D_D$ ($D_D < 0.1$) are denoted by blue squares and high-$D_D$ ($D_D > 0.1$) with red circles. Fireballs with $D_D$<0.008 have a yellow plus symbol. The normalized histogram for each component is stacked above the top frame of each panel. The number of bins was calculated based on Sturges's rule \citep{sturges1926choice}. Adapted from \citep{Hajdukova2024AA691A8H}.}
\label{fig:vel_components}
\end{figure*}

\begin{figure*}[h]
\centering
\includegraphics[width=1\textwidth]{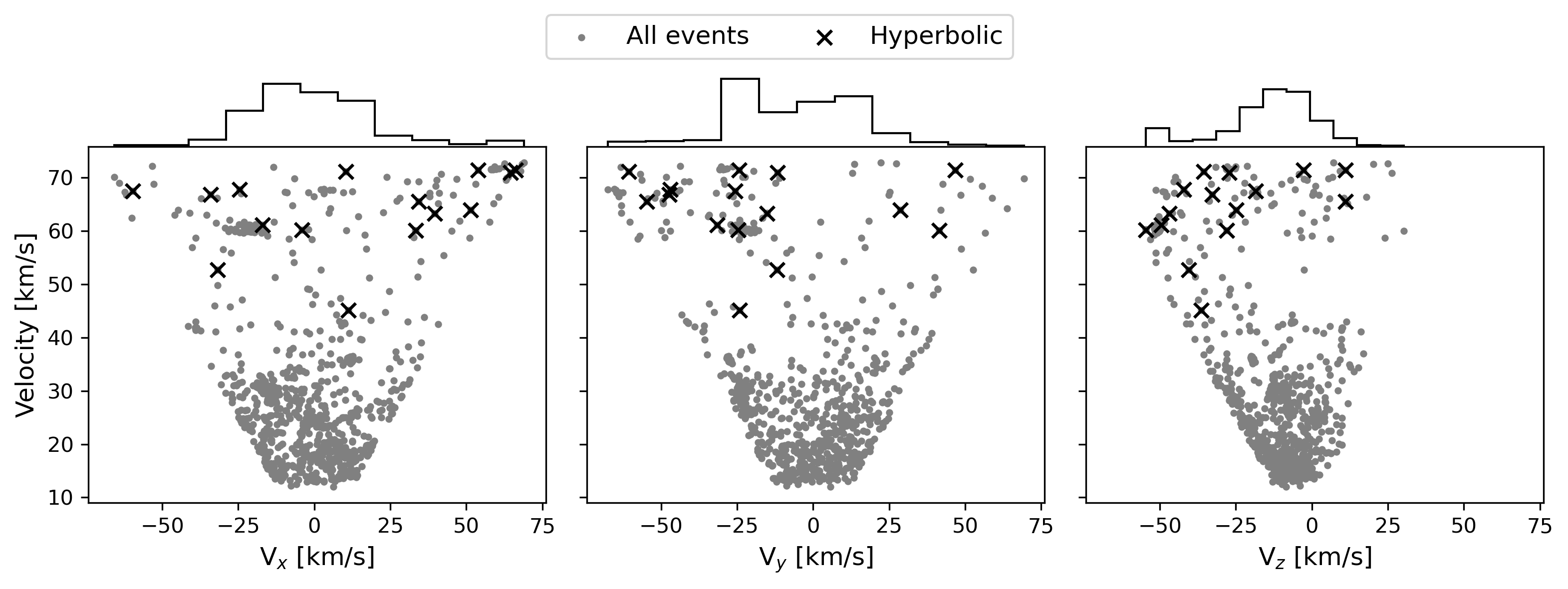}
\caption{Relationship between the total velocity and the pre-impact velocity components in a geocentric J2000 reference frame, as provided by the EN catalog. Hyperbolic fireballs are marked with crosses. The normalized histogram for each component is stacked above the top frame of each panel. The number of bins was calculated based on Sturges's rule \citep{sturges1926choice}. Same representation as in \citet{Hajdukova2024AA691A8H}.}
\label{fig:vel_components_EN}
\end{figure*}

Figure~\ref{fig:DD_vs_year_lat_energy_vel} illustrates the relationship between $D_D$ values and some key fireball parameters: time and total impact energy. Three fireballs that occurred during mid-2015 exhibit notably high $D_D$ values (top panel); the other high $D_D$ value event is the second oldest calibrated fireball. Moreover, Figure~\ref{fig:DD_vs_year_lat_energy_vel} demonstrates that events with higher $D_D$ values consistently exhibit lower than average energy levels (bottom panel). The full list of hyperbolic fireballs considered is provided in Table~\ref{tab:hyperbolics_data_e_i}, which includes their geolocation, energy, and pre-impact velocity components. Figures~\ref{fig:vel_components} and \ref{fig:vel_components_EN} show the distribution of these velocity components for the CNEOS and EN datasets.

\begin{table*}[h]
\centering
\caption{List of 6 hyperbolic fireballs in the CNEOS database.}
\begin{tabular}{lccccccccc}
\hline
Date (UTC) & Lat. [º] & Lon. [º] & Alt. [km] & E$_i$ [kt] & $V_x$ [km/s] & $V_y$ [km/s] & $V_z$ [km/s] \\
\hline
2022-07-28 01:36:08 & -6.0   & -86.9 & 37.5 & 0.68  & -17.1 & 23.5  & -7.2 \\
2021-05-06 05:54:27 & -34.7  & 141.0 & 31.0 & 0.076 & 9.6   & -24.4 & -4.6 \\
2017-03-09 04:16:37 & 40.5   & -18.0 & 23.0 & 1     & -15.3 & 25.8  & -20.8 \\
2015-02-17 13:19:50 & -8.0   & -11.2 & 39.0 & 0.11  & -28.2 & 3.4   & 4.6 \\
2014-01-08 17:05:34 & -1.3   & 147.6 & 18.7 & 0.11  & -3.4  & -43.5 & -10.3 \\
2009-04-10 18:42:45 & -44.7  & 25.7  & 32.4 & 0.73  & -18.9 & 2.6   & 0.3 \\
\hline
\end{tabular}
\tablefoot{Columns include the date of peak brightness, geodetic latitude and longitude, altitude, total impact energy, pre-impact velocity components in a geocentric Earth-fixed reference frame with axes aligned to Earth's rotation and equatorial plane, derived eccentricity, and orbital inclination. Bold values denote the velocity component with excess.}
\label{tab:hyperbolics_data_e_i}
\end{table*}

\end{appendix}

\end{document}